\documentclass[iop]{aastex}
\usepackage{color}
\usepackage{natbib}
\usepackage{lscape}

\newcommand{\AKARI}{\textit{AKARI}}

\def\Teff{$T_{\rm{eff}}$}
\def\Tcr{$T_{\rm{cr}}$}
\def\Tcond{$T_{\rm{cond}}$}
\def\logg{log \textit{g}}

\def\HtO{$\mathrm{H_{2}O}$}

\def\CHf{$\mathrm{CH_4}$}
\def\COt{$\mathrm{CO_2}$}

\def\AKARI{\textit{AKARI}}

\slugcomment{e-mail: satoko.sorahana@nao.ac.jp}

\shorttitle{Effect of Dust Size on the Near-Infrared Spectra (1.0--5.0~$\mu$m) of Brown Dwarf Atmospheres}
\shortauthors{Sorahana et al.}

\begin{document}


\title{Effect of Dust Size on the Near-Infrared Spectra (1.0--5.0~$\mu$m) of Brown Dwarf Atmospheres}


\author{Satoko Sorahana$^{1}$, Hiroshi Kobayashi$^{2}$, and Kyoko K. Tanaka$^{3}$}
\affil{$^{1}$National Opbservatory of Japan, Mitaka, Tokyo, 181-8588, Japan \linebreak
${^2}$Department of Physics, Nagoya University, Nagoya, Aichi 464-8602, Japan \linebreak
${^3}$Astronomical Institute, Tohoku University, Sendai, Miyagi 980-8578, Japan}
\email{satoko.sorahana@nao.ac.jp}


\begin{abstract}
In this study, we demonstrate the dependence of atmospheric dust size on the near-infrared spectra of ten L dwarfs, and constrain the sizes of dust grains in each L dwarf atmosphere. In previous studies, by comparing observed and modeled spectra, it was suggested that the deviations of their spectral shapes from theoretical prediction are general characteristics. Here, we focus on the dust size in brown dwarf atmospheres to understand the observed spectra. We confirm that changing the dust size changes the temperature-pressure structure of the atmosphere, with the shape of the spectrum changing accordingly. In the wavelength where dust is the main absorber of radiation (dust-dominated regime), a large dust opacity combined with a medium grain size, e.g. $0.1\,\mu$m results in a low photospheric temperature, and thus a small flux. Conversely, for the wavelength where gas absorption is dominant (gas-dominated regime),  a large dust opacity modifies the temperature-pressure structure, resulting in a high photospheric temperature, which corresponds to large flux emissions. Taking into account the size effect, we compare the model spectral fluxes in  the wavelength of 1--5~$\mu$m with the observational ones to constrain the main dust size in the atmosphere of each of the ten L dwarfs observed with $AKARI$ and SpeX or CGS4. Ultimately, we reveal that the observed data are reproduced with higher fidelity by models based on a medium dust size of 0.1--3.0~$\mu$m for six of these L dwarfs; therefore, we suggest that such atmospheric dust sizes apply to the majority of L dwarfs.
\end{abstract}





\section{Introduction}
In their earlier stages ($\sim$10$^6$ yrs), brown dwarfs are sustained by deuterium burning in their cores,  and simply cool off after this energy source is exhausted. As a consequence, thermonuclear processes do not dominate their evolution (\citealt{Burrows_2001}). 
Owing to the low temperature ({\Teff} $\sim$ 600--2200~K) and high density (log $P_{g}\sim6.0$ dyn/cm$^2$) of brown dwarf photospheres, they contain an abundance of molecules.
These photosphere conditions promote the condensation of atmospheric dust. This condensation process has been discussed since the 1960s (e.g. \citealt{Lord_1965,Larimer_1967a,Larimer_1967b}). 
The dust grains form below around 2000~K which is the condensation temperature of  refractory materials such as Fe, MgSiO$_3$, and Al$_2$O$_3$.
\citet{Helling_2001a} and \citet{Woitke_2003,Woitke_2004} considered the dust formation in the atmospheres of brown dwarfs to be similar to that around asymptotic giant branch (AGB) stars. \citet{Ackerman_2001}, \citet{Cooper_2003}, and \citet{Morley_2012} incorporated terrestrial cloud formation processes into the condensation process within brown dwarf atmospheres. \citet{Tsuji_1996a,Tsuji_1996b} and \citet{Allard_2001a} applied thermodynamic conditions to the dust formation. \citet{Tsuji_1996b,Tsuji_2002,Tsuji_2005} considered that dust forms at the equilibrium condensation temperature, {\Tcond}, assuming the local thermodynamic equilibrium (LTE) condition for the high-density gas in the photospheres of cool brown dwarfs.

Infrared spectroscopic observations are the most powerful tools for obtaining physical and chemical information concerning brown dwarf photospheres, including dust grains, as the majority of the energy emitted by brown dwarfs corresponds to infrared wavelength range, in which various molecular and dust features present. Dust grains in the photosphere contribute to the spectra directly via dust extinction. 
Furthermore, dust grains change the temperature structure of the photosphere indirectly owing to the effect of dust opacity.
Observations have revealed that the effect of dust grains appears most prominently in the near infrared spectra of L dwarfs (\citealt{Tsuji_1996b,Nakajima_2001}). By contrast, the spectra of T-dwarfs are less affected by dust grains. 
These results indicate that the influence of atmospheric dust is diminished around the L-T boundary (\citealt{Burgasser_2002,Burrows_2006,Morley_2012,Charnay_2018}).
One possible explanation for this is dust sedimentation within the atmosphere; if a dust cloud precipitate from the photosphere surface to deep inside the photosphere surrounding these spectral types, the effects of dust opacity in T-type spectra become smaller. However, the mechanism of dust disappearance has yet to be clarified. 

Current brown dwarf atmosphere models include a dust sedimentation effect as one of the model parameters, and such models can explain the observed spectral energy distribution (SED) relatively satisfactorily (e.g. \citealt{Tsuji_2002,Tsuji_2005,Saumon_2008}).
On the other hand, these models still struggle to explain several molecular absorption bands. \citet{Cushing_2008} reported that the model spectra, based on \citet{Marley_2002}, poorly fit the observed spectra in the 0.95--14.5~$\mu$m wavelength range for mid- to late-L dwarfs and early-T dwarfs. \citet{Cushing_2008} used data observed by the NASA Infrared Telescope Facility (IRTF; 0.9--2.5~$\mu$m and 3.0--4.0~$\mu$m) and the Spitzer Space Telescope (SST; 5.0--14.5~$\mu$m), concluding that the relatively poor fits corresponding to 3.0--4.0~$\mu$m range are most likely due to the limitations of their simple cloud model. 
In order to understand these limitations, it is important to analyze the 3.0--4.0~$\mu$m spectra of more objects.
\citet{Sorahana_2012} investigated the 2.5--5.0~$\mu$m spectra of sixteen objects observed by {\AKARI}, a Japanese infrared astronomical satellite, and found discrepancies between the molecular absorption bands in their spectra and the theoretical predictions for all objects. 
In our past studies, we considered vertical mixing and several non-equilibrium effects, e.g. chromospheric activity (\citealt{Sorahana_2014a}) and/or elemental abundances (\citealt{Sorahana_2014b}). However, those effects did not completely resolve the deviations between observations and models.
These deviations of the models may be caused by the treatment of clouds. 

In order to understand the deviations in the spectra between observation and model, 
we focused on the dust grain size of clouds in the brown dwarf atmospheres.
In previous unified cloudy models (UCMs), all grain sizes are 0.01\,$\mu$m (\citealt{Tsuji_2002}).
By contrast, we constructed a set of brown dwarf atmosphere models based on the UCM that adopt various dust grain sizes. 
We investigated the effects of different grain sizes on the atmospheric structure and infrared spectra from 1.0 to 5.0~$\mu$m, and constrain the main dust size in the atmosphere for each type of brown dwarf.

Our paper is organized as follows. In Section 2, we explain the original UCM, which is one of several brown dwarf atmospheric models, and the additional parameter of dust size $s$. In Section 3, we explain the observation data used in our analysis. Section 4 presents the effects of dust size on the atmospheric structure and infrared spectra. Then, in Section 5, we show the result of the main dust size for each object constrained by comparing the model spectra for each dust size model with observational spectra. Finally, the differences between the best-fit model and the other dust size models, as well as a possibility of dust size distribution, are discussed in Section 6.

\section{The Unified Cloudy Model}
The UCM is one of the brown dwarf atmosphere models that has been developed by Tsuji and colleagues (\citealt{Tsuji_1996a,Tsuji_1996b, Tsuji_2002, Tsuji_2005}). 
The physical parameters of general stellar model atmosphere include effective temperature {\Teff}, surface gravity {\logg}, elemental abundances, and micro turbulence velocity.
In addition to these parameters, the effect of dust incorporated into the UCM as follows. Dust forms at a layer in which the temperature falls to the extent that it is below the condensation temperature, {\Tcond} $\simeq2000$~K, according to the thermo-chemical equilibrium. The UCM considers dust particles composed of Fe, MgSiO$_3$, and Al$_2$O$_3$ \citep{Tsuji_2002}. The most important of these dust components is Fe, because it has the highest number density at the low temperatures encountered in brown dwarf atmospheres. 
At a high-altitude or in a low-temperature atmosphere, the size of dust grains may be so large that their sedimentation occurs quickly (\citealt{Burrows_1999}). 
The UCM introduced a critical temperature, {\Tcr}, below which the dust disappears.
The growth timescale of particles due to condensation is very sensitive to temperature. For the temperature $T>$ {\Tcr}, the growth of particles due to condensation is negligible compared to sedimentation. On the other hand, growth and sedimentation rapidly removes dust particles for $T<$ {\Tcr}. 
Therefore, the dust grains would exist only in the region for which {\Tcr} $< T < $ {\Tcond}.
Although {\Tcr} depends on the gas density and material composition, we simply treat it as a parameter. 
A lower {\Tcr} results in a thicker dust layer. 
The UCM is simpler than the other models, e.g. \citet{Ackerman_2001}. 
The simplicity of this model has an advantage that we can understand the dependence on new introduced parameters.
The temperature is determined under the assumption of an LTE in the calculation of the molecular abundances and dust formation. 
The chemical composition of all objects is assumed to be solar elemental abundances based on \citet{Allende_2002}.
The molecular-line absorption is calculated by the band model method during the construction of the model photosphere until the model has converged, and then detailed line lists are used to evaluate the final emergent spectra. The UCM applies the line lists of {\COt} (HITEMP database;
\citealt{Rothman_1997}), {\CHf} (\citealt{Freedman_2008} based on the Spherical Top Data System model of \citealt{Wenger_1998}), 
CO (\citealt{Guelachvili_1983, Chackerian_1983}), and {\HtO} \citep{Partridge_1997} (see also \citealt{Tsuji_2002} for details of the molecular line opacities). 
The other molecules/atoms and the corresponding line lists are described in \citet{Tsuji_2002}.
To determine the absorption line width, the micro turbulence velocity is set to be the solar value of 1~km/s. 

In the previous UCM model, the dust radius is fixed at 0.01\,$\mu$m. 
However, the dust size is determined by various processes (\citealt{Allard_2003}; \citealt{Ackerman_2001}; and \citealt{Helling_2008c}), and the mean dust size has a large uncertainty due  to the complexity of the processes. 
In this study, we thus set the dust radius $s$ as a new calculation parameter. 
We investigate the spectral response for seven different radii: $s = 0.01$, 0.03, 0.1, 0.3, 1, 3, and 10\,$\mu$m, clarify how the size of dust grains affects the infrared spectrum, and constrain the size of the dust in the atmosphere for each object from its spectrum. 
For each case, we calculated a total of 153 models (parameter grids: $1700$K$<${\Tcr}$<1900$K, $4.5<${\logg}$<5.5$, and $1000$K$<${\Teff}$<2600$K), taking into account the radiative transfer based on the gas and dust opacities.
From a total of 1071 models 20 are diverged (Table~\ref{unconvergenttlist}) and were excluded from the subsequent analyses.

\begin{deluxetable}{llll}
\tabletypesize{\scriptsize}
 \tablecaption{Twenty Models without Convergent \label{unconvergenttlist}}
\tablewidth{0pt}
\tablehead{
\colhead{{\Tcr}}&\colhead{{\logg}}  &\colhead{{\Teff}} &\colhead{dust size} \\
\colhead{[K]}&\colhead{}  &\colhead{[K]} &\colhead{[$\mu$m]} }
\startdata
1700&4.5&1900&0.03\\
1700&5.0&1900&0.03\\
1700&5.5&1900&0.03\\
1700&5.5&2000&0.03\\
1700&5.5&2100&0.03\\
1800&4.5&1900&0.03\\
1800&5.0&2200&0.03\\
1900&5.0&1900&0.03\\
1700&4.5&2100&0.1\\
1800&5.0&1900&0.1\\
1900&5.0&1700&0.1\\
1900&5.5&2000&0.1\\
1900&5.5&2100&0.1\\
1700&4.5&2100&0.3\\
1700&4.5&2200&0.3\\
1700&5.5&2000&0.3\\
1800&5.5&2200&0.3\\
1800&5.5&2300&0.3\\
1900&5.5&1900&1\\
1700&5.5&1900&3\\
\enddata
\end{deluxetable}

\section{Observed Data for Our Analysis} 
We attempted to constrain the main size of the dust grains in the atmosphere for each brown dwarf by comparing spectra observed by {\AKARI} and SpeX or CGS4 with 1051 model spectra predicted by the UCM.
We focus on only L dwarfs 
because their spectra are affected by dust more strongly than those of T dwarfs.
We also remove the L9, which corresponds to the L/T transition object, from the comparison list, because it is known that, among brown dwarfs, many L/T transition objects possess particularly complex atmospheric structures. 
The list of the 10 objects analyzed in this study are shown in Table~\ref{objectlist}.

\subsection{{\AKARI} Data}
{\AKARI}, the Japanese infrared astronomical satellite, has an infrared reflecting telescope, with an aperture of 68.5~cm, and two scientific instruments: the Far-Infrared Surveyor (FIS; \citealt{Kawada_2007}) and the InfraRed Camera (IRC; \citealt{Onaka_2007}). The FIS and IRC cover wavelengths of 50--180~$\mu$m and 1.8--26~$\mu$m, respectively. The liquid-He-cooled holding period of observations (Phase 1, 2) lasted from May 2006 until August 2007. After the boil-off of liquid-He, cryocooler-assisted observations were continued using only the near-infrared (NIR) camera of the IRC (Phase 3). 
The data used in this study were obtained using the IRC NIR channel (2.5--5.0~$\mu$m) in grism mode ($R=\lambda / \Delta \lambda \sim 120$). Precise details of the observational data are provided elsewhere
\citet{Sorahana_2012}.

\subsection{SpeX Data}
SpeX is a medium-resolution spectrograph covering 0.8--5.4~$\mu$m onboard the NASA IRTF on Mauna Kea, Hawaii. The IRTF is a 3.0~m telescope optimized for infrared observations. A high throughput prism mode that uses a single order long slit (60 arcsec) provides a spectral resolution of $\lambda / \Delta \lambda =R \sim 100$ for 0.8--2.5~$\mu$m. Using prism cross-dispersers (for 15-arcsec-long slits), $R$ becomes 1000--2000 across 0.8--2.4~$\mu$m.
Nine of the L dwarfs in our sample set have been observed by \citet{Burgasser_2004,Burgasser_2006a,Burgasser_2008,Burgasser_2010}, \citet{Burgasser_2007}, and \citet{Cushing_2004} with SpeX. The data sets were obtained using its low-resolution prism-dispersed mode with resolutions of 75--200 depending on the used slit width. For these objects, we retrieved the data from the SpeX Prism Spectral Libraries built by Adam Burgasser. 
We used spectral data for wavelengths $\lambda$ longer than 1.0~$\mu$m. 

\subsection{CGS4 Data}
Only SDSS~J1446+0024 was unpublished, with the data obtained from D. Looper (2010, private communication). The data was obtained using the CGS4 cassegrain instrument on board the United Kingdom Infrared Telescope (UKIRT), which is a 3.8~m infrared reflecting telescope located on Mauna Kea, Hawaii.
CGS4 is a grating spectrometer with a spectral resolution $R$ between 1,000 and 30,000.
SDSS~J1446+0024 was observed at $R \sim$1000.
As this data did not include errors, we assigned 10\% of each flux as the flux error of each data point. 
We use the CGS4 data in the wavelength range longer than 1.0~$\mu$m. 

\begin{deluxetable}{llcrcc}
\tabletypesize{\scriptsize}
\tablecaption{Ten Brown Dwarfs Observed by {\AKARI} \label{objectlist}}
\tablewidth{0pt}
\tablehead{
\colhead{No. }&\colhead{Object Name}  &\colhead{Object Name in This Paper}&\colhead{Sp. Type}   &  \colhead{} &  \colhead{References} 
}
\startdata
1&2MASS~J14392836+1929149&2MASS~J1439+1929&L1 &&1\\
2&2MASS~J00361617+1821104 &2MASS~J0036+1821& L4 &&2\\
3&2MASS~J22244381--0158521 &2MASS~J2224--0158& L4.5 &&1\\
4&SDSS~J053951.99--005902.0 &SDSS~J0539--0059&L5   &&2\\
5&SDSS~J144600.60+002452.0  &SDSS~J1446+0024&L5    &&2\\
6&2MASS~J15074769--1627386&2MASS~J1507--1627&L5  &&1\\
7&GJ~1001B &GJ~1001B& L5 &&1\\
8&2MASS~J08251968+2115521 &2MASS~J0825+2115& L6 &&2\\
9&2MASS~J16322911+1904407 &2MASS~J1632+1904&L7.5  &&2\\
10&2MASS~J15232263+3014562 &2MASS~J1523+3014& L8 &&2\\
\enddata
\vspace{-5mm}
\tablecomments{Reference of spectral type (1) \citet{Kirkpatrick_2000}; (2) \citet{Geballe_2002}.} 
\end{deluxetable}

\section{Effects on Atmospheric Structure and Infrared Spectra}
\subsection{Dust size dependence}
\begin{figure}[!t]
\begin{center}
\plotone{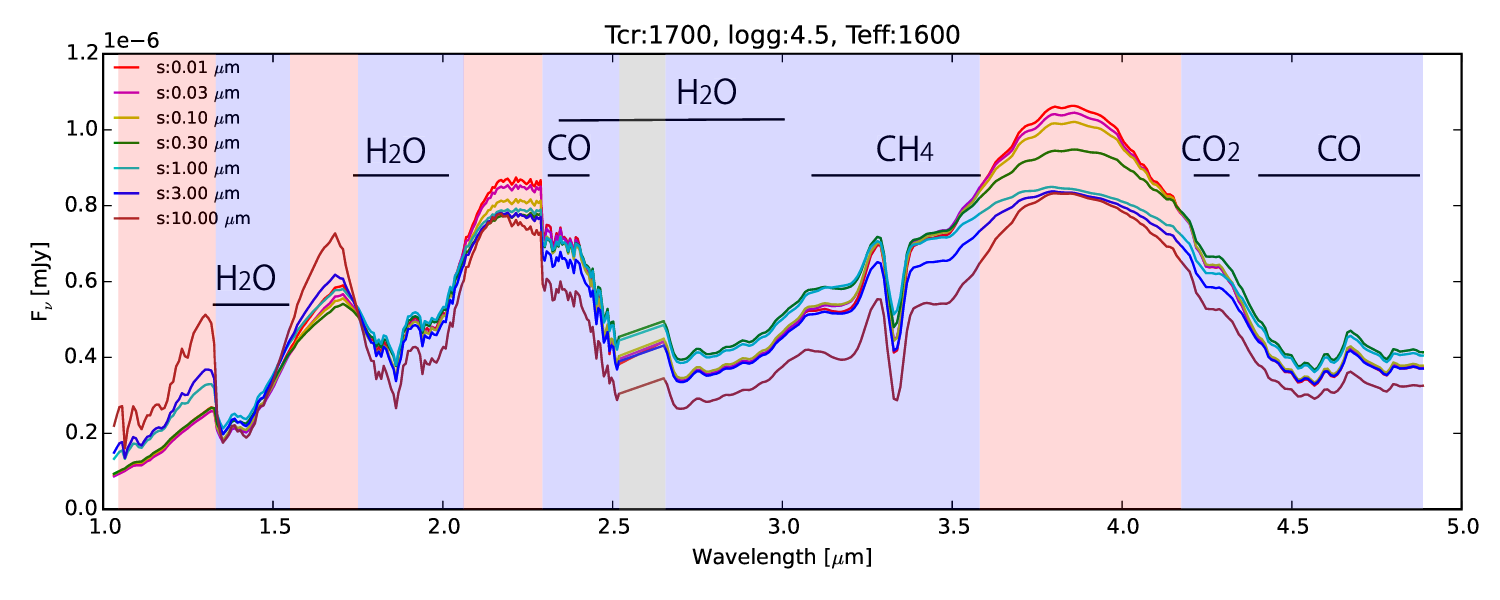}
\end{center}
\caption{Comparison of seven models with different dust sizes ($s =0.01$, 0.03, 0.1, 0.3, 1, 3, and 10~$\mu$m) for the parameter set of $(${\Tcr}/{\logg}/{\Teff}$)=(1700$K$/4.5/1600$K) as an example. Major molecular absorption bands, {\HtO}, CO, {\COt}, and {\CHf} are shown.
The gray area indicates an absence of data.
The red and blue areas indicates the dust- and gas-dominated regimes, respectively.
} 
\label{f1}
\end{figure}

First, we compared seven atmospheric models with different dust sizes ($s =0.01$, 0.03, 0.1, 0.3, 1, 3, and 10~$\mu$m).
Figure \ref{f1} shows the model comparison for a typical L dwarf with $(${\Tcr}/{\logg}/{\Teff}$)=(1700$K$/4.5/1600$K) as an example.
This comparative analysis reveals that the shape of near-infrared spectra does depend on dust size in the objects' atmospheres.
The spectral shapes relating to dust sizes of $s= 0.01\mu$m and 0.03~$\mu$m show minimal differences. 
For large dust sizes (e.g. $s \ga 0.1$~$\mu$m) by contrast, the entire spectral shapes change significantly.
In addition, the effect of dust size on the spectral shapes depends on the wavelength $\lambda$.
As shown in Figure~\ref{f1}, gases (H$_2$O, CH$_4$, CO, CO$_2$) are dominant absorbers in the ranges $\lambda =$ 1.3--1.5, 1.8--2.1, 2.4--3.5, \& 4.3--5.0~$\mu$m, which we designate as the gas-dominated regimes. 
Conversely, dust is more important in the other wavelength bands ($\lambda =$ 0.8--1.3, 1.5--1.8, 2.1--2.4, \& 3.5--4.3~$\mu$m), which we designate as the dust-dominated regimes. 
The dust size dependence on the fluxes in dust- or gas-dominated regions is different (\citealt{Burrows_2006}). We below clarify the size dependence according to the detailed analysis of dust opacities.

\begin{figure}[!]
\epsscale{.8} 
\begin{center}
\plotone{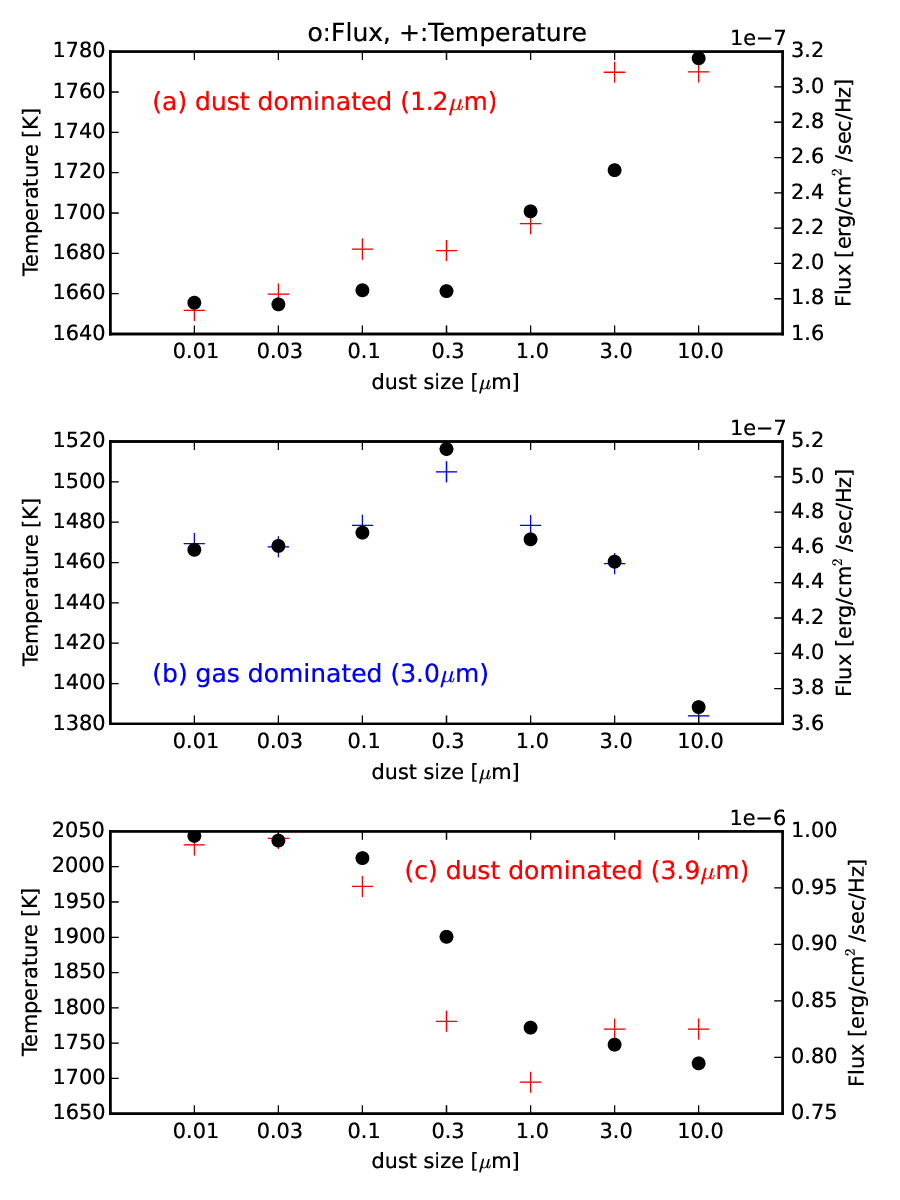}
\end{center}
\caption{Temperature (crosses) and flux (circles) for the dust- (red color: $\lambda=$1.2 and 3.9~$\mu$m) and gas-dominated regimes (blue color: $\lambda=$3.0~$\mu$m) of the photosphere at $\tau_\lambda \approx 1$. } \label{f0}
\end{figure}

The fluxes at certain wavelengths are related to the temperatures of the photospheres for these wavelengths (i.e., optical depth at a specific wavelength $\tau_\lambda \approx 1$). 
Figure~\ref{f0} shows the temperature and flux at $\tau_\lambda = 1$ both for the dust-dominated ($\lambda = 1.2$ and 3.9~$\mu$m) and gas-dominated ($\lambda = 3.0$~$\mu$m) regimes 
as a function of dust grain size. 
Almost all flux values increase with the photospheric temperatures at $\tau_\lambda\sim1$, which are independent of both dust size and the two regimes.
This is because the flux is determined approximately by the blackbody radiation of the photosphere at $\tau_\lambda\sim1$. 

\begin{figure}[!]
\epsscale{.8} 
\begin{center}
\plotone{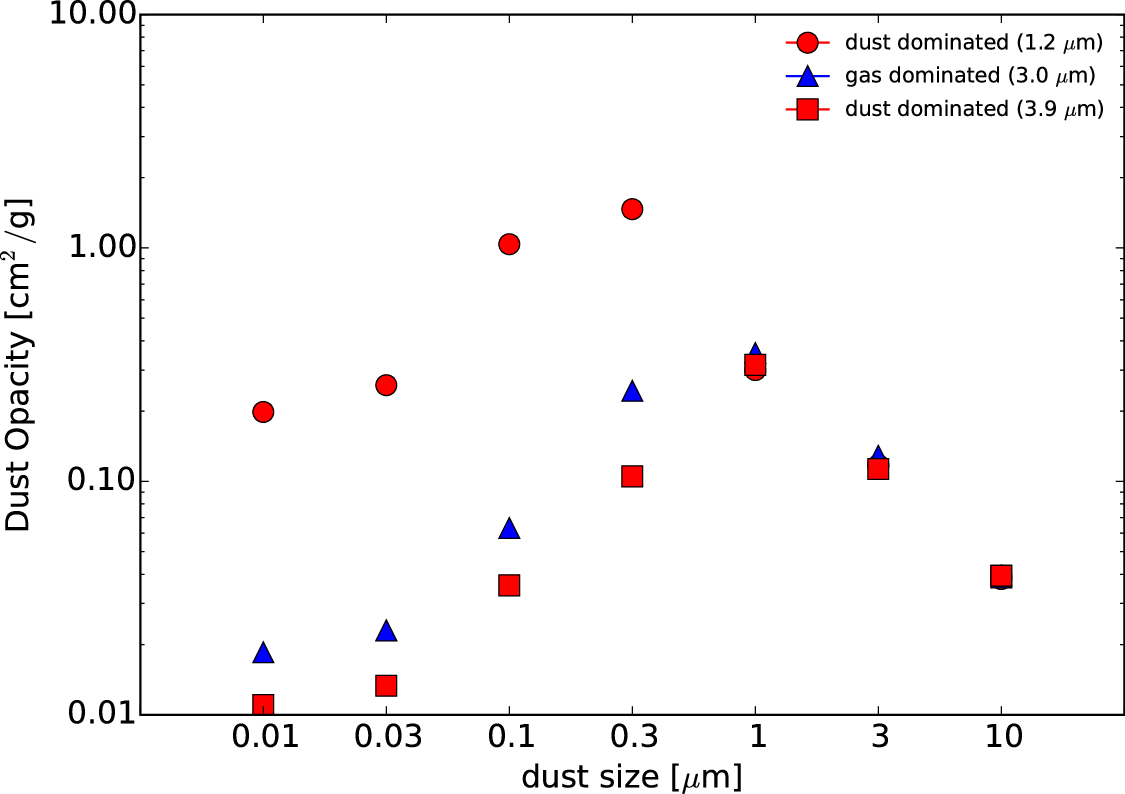}
\end{center}
\caption{Dust opacity (including iron, enstatite, and corundum) at wavelengths of 1.2 (circles), 3.0 (triangles), and 3.9 (squares)~$\mu$m. Red and blue-colored data correspond to the dust- and gas-dominated regimes in Figure~\ref{f0}, respectively.
} 
\label{qext}
\end{figure}

Figure~\ref{qext} shows dust opacities as a function of dust size.
The dust opacity changes as the dust size in the atmosphere changes.
At wavelengths shorter than 2.5~$\mu$m, medium-sized dust grains of about 0.3~$\mu$m have the largest opacity, whereas grain sizes from 3--10~$\mu$m lead to the smallest opacity. Similarly, at wavelengths longer than 2.5~$\mu$m, medium-sized dust measuring 1~$\mu$m has the largest opacity; however, small opacities are realized for smaller and larger grain sizes alike. Figure~\ref{qext} illustrates this trend within three wavelength regions, including both dust- and gas-dominated regions. 

\begin{figure}[!]
\epsscale{.8} 
\begin{center}
\plotone{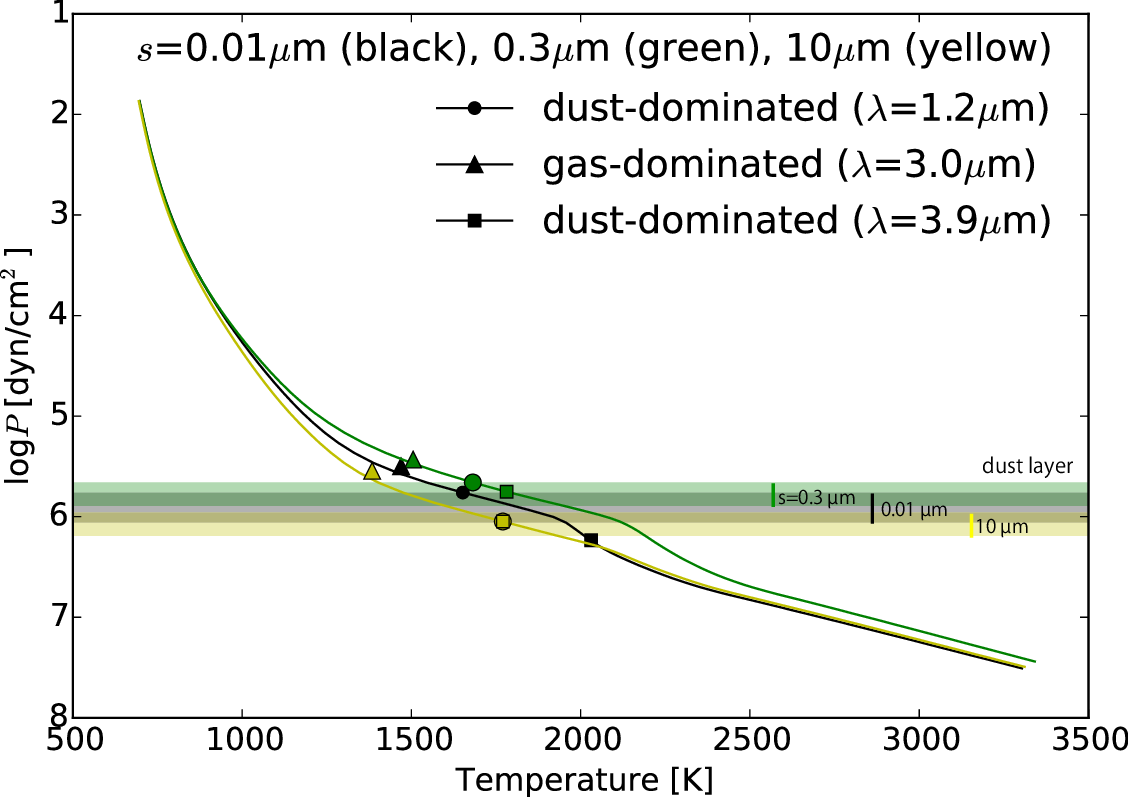}
\end{center}
\caption{Temperature structure for the models with each dust size $s$ of 0.01 (black), 0.3 (green), and 10~$\mu$m (yellow).
The other three parameters, {\Tcr}, {\logg}, and {\Teff} are 1700~K, 4.5, and 1600~K, respectively.
The photospheric surface corresponding to $\tau\sim1$ at $\lambda$= 1.2 (dust-dominated regime), 3.0 (gas-dominated regime), and 3.9~$\mu$m (dust-dominated regime) are shown in circles, triangles, and squares, respectively. Dust layer for each model is also shown in the fill-in.
} 
\label{tempstr}
\end{figure}

For the dust-dominated regime, 
the photosphere exists around the dust layer as shown in Figure~\ref{tempstr}.
Therefore, the photospheric temperature directly depends on the dust opacity.
If the opacity is very large ($\ga 2\times10^{-1} \,{\rm cm}^2/{\rm g}$ in Figure~\ref{qext}), 
then the photosphere at $\tau_\lambda\sim1$ is around the top of dust cloud and the photospheric temperature is close to the temperature at the top of the cloud, $T_{\rm cr}=1700$\,K, as observed for dust sizes of $ s \la 1.0 \,\micron$ and $\lambda = 1.2\,\micron$ (Figure~\ref{f0}a) or $s\sim1.0\,\micron$ and $\lambda=3.9\,\micron$ (Figure~\ref{f0}c). 
For low opacities ($< 2\times10^{-1} \,{\rm cm}^2/{\rm g}$ in Figure~\ref{qext}), the photosphere at $\tau_\lambda\sim1$ overlaps with the dust cloud with the result that the photospheric temperature increases with decreasing opacity. 
This behavior can be observed for $s \ga 3\,\micron$ and $\lambda = 1.2\,\micron$ (Figure~\ref{f0}a) or for $s\la0.3\,\micron$ and $s\ga 3.0\,\micron$ in the case of $\lambda = 3.9\,\micron$ (Figure~\ref{f0}c). 

\begin{figure}[!]
\epsscale{.8} 
\begin{center}
\plotone{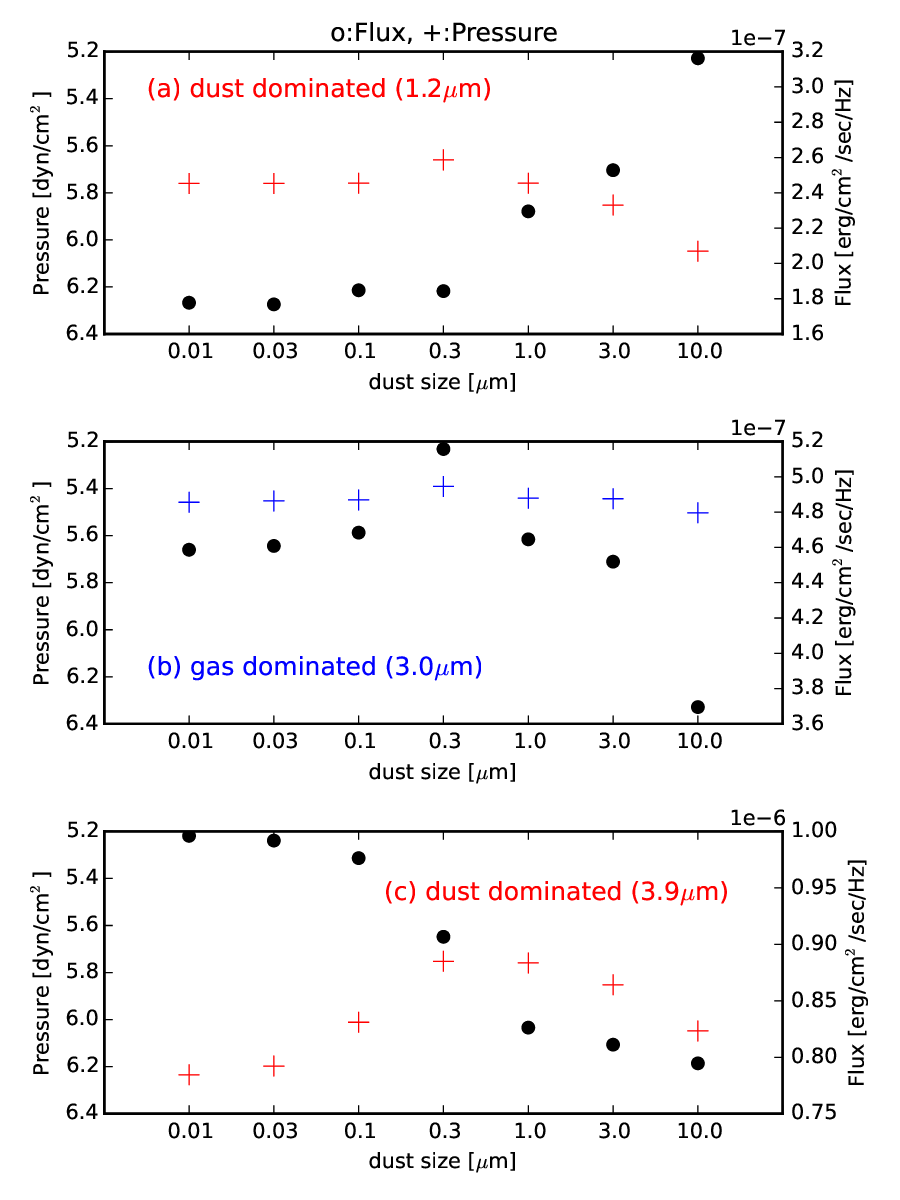}
\end{center}
\caption{Pressure (crosses) and flux (circles) for the dust (red color: $\lambda=$1.2 and 3.9~$\mu$m) and gas-dominated regimes (blue color: $\lambda=$3.0~$\mu$m) of the photosphere at $\tau_\lambda \approx 1$. 
} 
\label{PvsDust}
\end{figure}

Contrary to the results for the dust-dominated regime, in the gas-dominated regime, the opacity of the dust grains affects the flux value indirectly.
Because column mass density above the photosphere is determined chiefly by the gas opacity, the pressure within the photosphere is independent of dust grain size as shown in Figure~\ref{PvsDust}(b). 
On the other hand, the temperature of the photosphere is dependent on the grain size because the temperature structure is determined by the dust opacity. 
Figure~\ref{tempstr} shows the temperature structures for the models with each small ($s=0.01~\mu$m), medium ($s=0.3~\mu$m), and large ($s=10~\mu$m) dust size. 
The photospheric surface ($\tau_\lambda\sim1$) for gas-dominated regime of each model, shown in triangle in Figure~\ref{tempstr}, is located above each dust layer and has different photospheric surface temperatures for each model.
If the total dust opacity integrated at each wavelength is large, e.g. $s=0.3\,\micron$, the temperature does not decrease easily before reaching the photosphere (green line in Figure~\ref{tempstr}).
Therefore, a large dust opacity results in a high photospheric temperature, which leads to large flux emissions. 
By contrast, if the total dust opacity integrated at each wavelength is low, e.g. $s=10\,\micron$, the temperature of the photosphere is low because the warming effect of the dust is almost eliminated enabling the temperature to decrease easily (yellow line in Figure~\ref{tempstr}).

Some previous studies suggest that the L--T transition, photometry, and spectra of brown dwarfs are explained without dust clouds under the assumption of shallow temperature gradients caused by fingering convection and thermochemical instabilities of some molecular abundances (\citealt{Tremblin_2016}). This effect makes the temperatures at the photospheres weakly dependent on wavelength, which is similar to those due to dust cloud in the dust-dominated wavelengths if $s \ga 1\micron$. On the other hand, the photospheric temperatures determined by dust clouds are relatively higher at {\AKARI} wavelengths because of lower opacities for $s \la 1\micron$ (see Figs.~3 and~4). Therefore, dust clouds are more responsible for the spectra of brown dwarfs likely having $s \la 1 \micron$.

\subsection{Comparison with observations}
\begin{figure}[!]
\epsscale{1} 
\begin{center}
\plotone{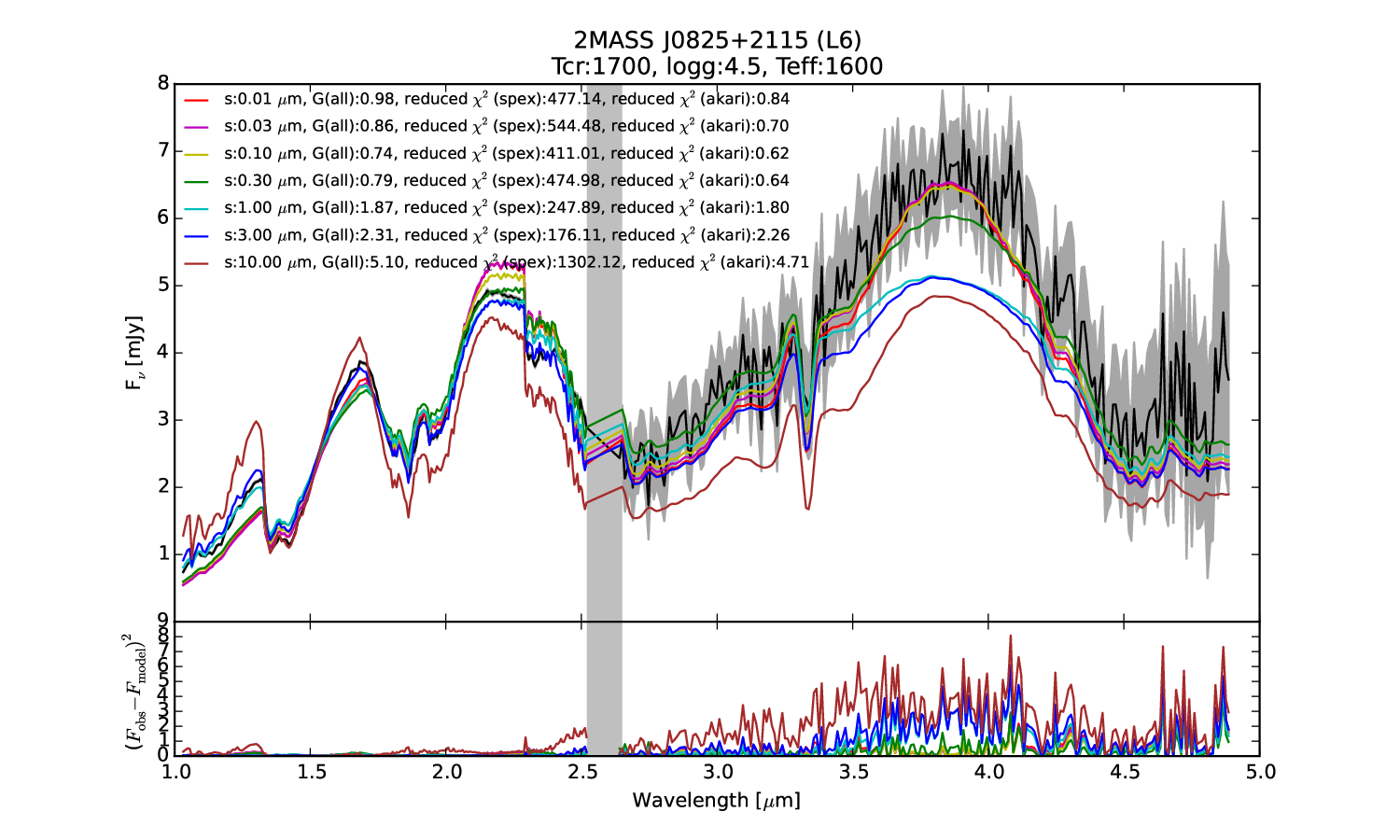}
\end{center}
\caption{Comparison between observed (black), observed with errors (grey), and modeled (color) spectra for 2MASS~J0825+2115. Models with different dust sizes are color-coded. 
Bottom panel shows the difference between observation and the new best-fit model (red) or the 0.01~$\mu$m best-fit model (blue).} 
\label{f9}
\end{figure}

We compared the spectra of 2MASS~J0825+2115 observed by {\AKARI} and SpeX with model spectra for different dust grain sizes.
The fitting method is described in detail in Section 5.1.
Object 2MASS~J0825+2115 is a typical L dwarf (spectral type L6). 
We investigated the dust size dependence of the model of ({\Tcr}/{\logg}/{\Teff})=(1700/4.5/1600) against 2MASS~J0825+2115 (see Figure~\ref{f9}).
The results demonstrate that the atmospheric model corresponding to a dust size of 0.1~$\mu$m is able to reproduce the observed spectrum with greater fidelity than atmospheric models based on other dust sizes.
The modeled spectrum representing the original UCM with a dust size of $s = 0.01$~$\mu$m generally reproduces the observed spectrum; however, not for the entire wavelength range.
Irrespective of this, our simulated spectrum for $s=0.1$~$\mu$m outperforms the original UCM for $s =0.01$~$\mu$m.
In particular, the flux values relating to $\lambda = 3.0$--3.5~$\mu$m, and 2.1--2.3~$\mu$m are improved (yellow and red for $s = 0.1$ and 0.01~$\mu$m, respectively in Figure~\ref{f9} and see also Figure~\ref{figall}h).
For $\lambda = 3.0$--3.5~$\mu$m, the main reason for this is the dust size dependency in the gas-dominated regime. Here, the photospheric temperature increases due to the large dust opacity, resulting in large fluxes for a large dust size of $s=0.1\, \micron$. 
Alternatively, for the dust-dominated regime $\lambda = 2.1$--2.3\,\micron, 
the photospheric temperature decreases due to the large dust opacity, resulting in small fluxes for large dust grains with $s=0.1\, \micron$.
Considering both the gas- and the dust-dominated regimes, 
a simulated spectrum based on a dust size of $s \approx 0.1\,\micron$ is likely to reproduce the observed fluxes of 2MASS~J0825+2115. 

\section{Constraints on the Dust Size of Each Object}

\subsection{Method of Derivation of Best-fit Model for Each Object}

\citet{Sorahana_2012} tried to produce models explaining both the {\AKARI} and SpeX or CGS4 (hereafter SpeX/CGS4) observed spectra with reasonable success. 
However, almost all of these models did not provide an accurate fit spanning the entire of 1--5~$\mu$m wavelength range, and there were always some deviations either in the {\AKARI} or the SpeX/CGS4 wavelength. 
This problem of imperfect broadband spectral agreement was also reported by \citet{Cushing_2008}. 
In this paper, we have introduced a new parameter of dust size $s$, which helps the models to reproduce the observed spectra. 
Despite this, it remains difficult to derive the best-fit model by basic $\chi^2$ fitting because we used the different data sets of {\AKARI} and SpeX/CGS4.
As the SpeX errors are extremely small and the number of data is large compared to the {\AKARI} data, 
the SpeX data has a greater weighting than the {\AKARI} data in a simple $\chi^2$ fitting, 
and thus the $\chi^2$-fitted model does not reproduce the {\AKARI} spectra.
Therefore, we used a goodness of fit parameter, $G_k$, defined by the following equation, which treats the SpeX and {\AKARI} data almost equally:
\begin{equation}
\label{G}
G_{k} = w_{\mathrm{spex}}  \chi^2_{\mathrm{spex},k} + w_{\mathrm{akari}}  \chi^2_{\mathrm{akari},k} ,
\end{equation}
where $\chi^2_{\mathrm{spex},k}$ and $\chi^2_{\mathrm{akari},k}$ are the reduced $\chi^2$ of each SpeX and {\AKARI} data point given as
\begin{equation}
\label{chi2}
\chi^2_{k} = \frac{1}{N}{\displaystyle \sum_i} \left( \frac{f_{i} - C_{k}F_{k,i}}{\sigma_{i}} \right)^2,
\end{equation}
where $N$ is the degree of freedom, i.e., the number of parameters subtracted from the number of observational data points,
$f_{i}$ and $F_{k, i}$ are the flux densities of the $i$-th observed data point and $k$-th model, respectively, 
the variance ${\sigma_{i}}$ denotes the error associated with the $i$-th observed data point, 
and $C_{k}$ is the scaling factor given by 
\begin{equation}
\label{scalingfactor}
C_{k} = \frac{\sum f_{i} F_{k,i}/\sigma_{i}^2}{\sum {F_{k,i}}^2/{\sigma_{i}^2}}.
\end{equation}
We introduced the scaling factor, to find the best-fit model considering the uncertainty in the distance to the object, the dust extinction of the object, etc. (see also \citealt{Cushing_2008} for further details).
$w_\mathrm{spex}$ and $w_\mathrm{akari}$ in Equation~(\ref{G}) are the weighting coefficients applied to the SpeX and {\AKARI} data, respectively, and are expressed as
\begin{equation}
\label{w}
w_\mathrm{spex} = \frac{x_\mathrm{akari}}{x_\mathrm{spex} + x_\mathrm{akari}}, \hspace{5mm}
w_\mathrm{akari} = \frac{x_\mathrm{spex}}{x_\mathrm{spex} + x_\mathrm{akari}}.
\end{equation}
The sum of $w_\mathrm{spex}$ and $w_\mathrm{akari}$ equals 1. Each $x_\mathrm{spex}$ and $x_\mathrm{akari}$ are given by
\begin{equation}
\label{x}
x_\mathrm{spex} = \frac{1}{N_\mathrm{spex}} {\displaystyle \sum_i} \frac{1}{\sigma_{\mathrm{spex},i}^2}, \hspace{5mm}
x_\mathrm{akari} = \frac{1}{N_\mathrm{akari}} {\displaystyle \sum_i} \frac{1}{\sigma_{\mathrm{akari},i}^2},
\end{equation}
where, $N_\mathrm{spex}$ and $N_\mathrm{akari}$ are the degree of freedom for SpeX and {\AKARI}, respectively, and $\sigma_{\mathrm{spex},i}$ and $\sigma_{\mathrm{akari},i}$ are errors for the $i$-th observed data point in the SpeX and {\AKARI} data, respectively.
The smaller the value obtained for $G_k$, the better the fit. 
Furthermore, we applied this method to SDSS~J1446+0024, which uses the CGS4 data at shorter wavelengths.

\subsection{Dust Size of the Best-fit Model for Each Object}
We derived the best-fit models for 10 objects, as shown in Table~\ref{modeltable}.
We chose the model with minimum $G_k$ value as the best-fit model.
Each best-fit model corresponded to a medium dust grain size (0.1--3.0~$\mu$m), except for SDSS~J2224--0158 (L4.5) and SDSS~J1446+0024 (L5).
For these two objects, smaller dust size models (0.01~$\mu$m) produced the best-fits.

\begin{figure}
\epsscale{1} 
\begin{center}
\plotone{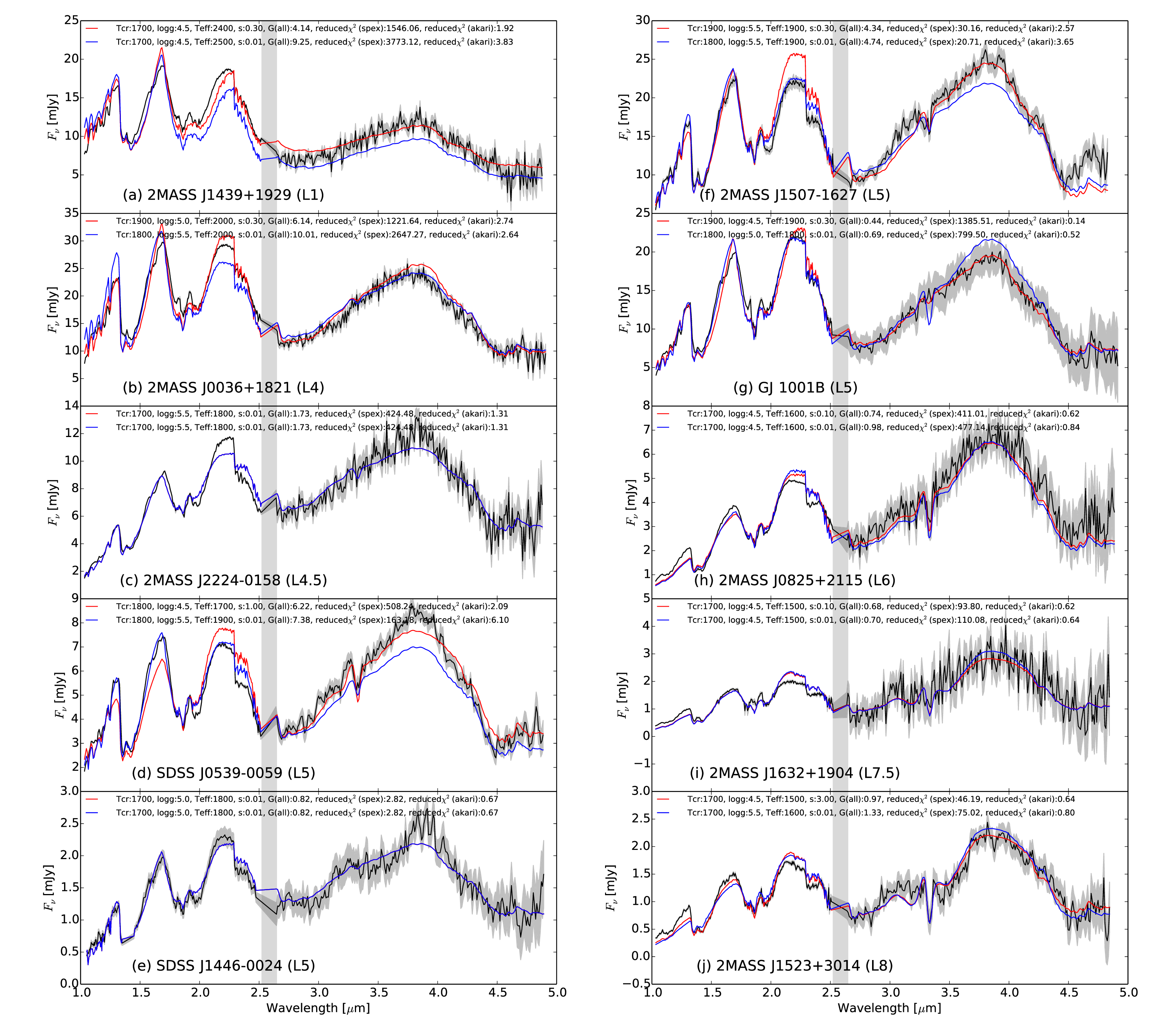}
\end{center}
\vspace{-1cm}
\caption{Comparison between the observed spectrum and two modeled spectra for each object. For the modeled spectra, one is the model based on a dust grain size of 0.01~$\mu$m (blue), and the other is the new best-fit model deduced after considering seven dust grain sizes (red).} 
\label{figall}
\end{figure}

In addition, Figure~\ref{figall} shows the individual results of a comparison between two models and the observation of each object; one model is a best-fit model derived using only the models with a dust size of 0.01~$\mu$m (blue in Figure~\ref{figall}; used in the discussion in Section~6.1), while the other model is a new best-fit model derived by considering the models based on the seven dust sizes (red in Figure~\ref{figall}).

For 2MASS~J1439+1929 (L1), the dust size of the best-fit model is 0.3$\, \micron$.
The theoretical spectrum in the {\AKARI} region reproduces the observed spectrum relatively well. 
In the SpeX regime, the $J$ and $K$ bands demonstrate similar fitting accuracy, although the head of the $H$ band does not fit well.
Similarly, for 2MASS~J0036+1821 (L4), the dust size of the best-fit model is 0.3$\, \micron$.
The theoretical spectrum in the {\AKARI} region also fits the observed spectra relatively well. In this case, although the $J$ band is relatively well fitted, the head of the $H$ and $K$ bands are not well fitted in the SpeX regime.
For 2MASS~J2224-0158 (L4.5), the dust size of the best-fit model is 0.01$\, \micron$.
Although the $K$ band of the simulated spectrum does not fit with the observed counterpart,
the simulated and observed spectra otherwise agree relatively closely.
For SDSS~J0539--0059 (L5), the dust size of the best-fit model is 1.0$\, \micron$; however, the observations and models do not fit well across the entire wavelength range.
The {\AKARI} region is consistent with the modeled and observed spectra except for the flux peak at $\sim 3.8\, \micron$, while the shape of the spectrum in the SpeX region is not consistent overall.
For SDSS~J1446+0024 (L5), the dust size of the best-fit model is 0.01$\, \micron$.
In this case, the model can explain the entire observed spectrum, though {\AKARI} data has poor signal-to-noise ratio.
For 2MASS~J1507--1627 (L5), the dust size of the best-fit model is 0.3$\, \micron$.
Only the $K$ band does not fit between the modeled and observed spectra.
For GJ~1001B, the dust size of the best-fit model is 0.3$\, \micron$.
The model and the observed spectra demonstrate a close agreement overall, but there are slight deviations in the $H$ and $K$ bands.
For 2MASS~J0825+2115 (L6), the dust size of the best-fit model is 0.1$\, \micron$.
Here, the simulated and observed spectra are relatively consistent for the {\AKARI} region; however, the slope of the spectrum in the SpeX region does not demonstrate satisfactory consistency.
For 2MASS-J1632+1904 (L7.5) and 2MASS~J1523+3014 (L8), the dust sizes of the best-fit models are 0.3 and 3.0$\, \micron$, respectively.
Both these objects exhibit similar model-observation agreements as revealed for 2MASS~J0825+2115.

\begin{landscape}
\begin{deluxetable}{llcrllllllrcc}
\tabletypesize{\scriptsize}
\tablecaption{Physical Parameters of the Best-Fit Models for Ten Brown Dwarfs \label{modeltable}}
\tablewidth{0pt}
\tablehead{
\colhead{No. }&\colhead{Object Name}  &\colhead{Sp. Type} &\multicolumn{3}{c}{0.01~$\mu$m}  &&\multicolumn{4}{c}{New}  &\colhead{$\chi^2_{\mathrm{spex},\mathrm{new}}$-$\chi^2_{\mathrm{spex},\mathrm{0.01{\mu}\mathrm{m}}}$}  &\colhead{$\chi^2_{\mathrm{akari},\mathrm{new}}$-$\chi^2_{\mathrm{akari},\mathrm{0.01{\mu}\mathrm{m}}}$} \\ 
\cline{4-6} \cline{8-11} 
&&&\colhead{{\Tcr}}&\colhead{{\logg}}  &\colhead{{\Teff}} &&\colhead{{\Tcr}}&\colhead{{\logg}}  &\colhead{{\Teff}} &\colhead{dust size}  
}
\startdata
1&2MASS~J1439+1929&L1 &1700&4.5&2500&&1700& 4.5& 2400& 0.3&$-$&$-$\\
2&2MASS~J0036+1821& L4 &1800&5.5&2000&&1900&5.0&2000&0.3&$-$&$+$\\
3&2MASS~J2224--0158& L4.5 &1700&5.5&1800&&1700& 5.5&1800&0.01&0&0\\
4&SDSS~J0539--0059&L5   &1800&5.5&1900&&1800&4.5& 1700&1.0&$+$&$-$\\
5&SDSS~J1446+0024&L5    &1700&5.0&1800&&1700& 5.0& 1800& 0.01&0&0\\
6&2MASS~J1507--1627&L5  &1800&5.5&1900&&1800&5.0& 1700&1.0&$+$&$-$\\
7&GJ~1001B& L5 &1800&5.5&1900&&1900&4.5& 1900& 0.3&$-$&$-$\\
8&2MASS~J0825+2115& L6 &1700&4.5&1600&&1700& 4.5& 1600& 0.1&$-$&$-$\\
9&2MASS~J1632+1904&L7.5 &1700&4.5&1500 &&1700&4.5& 1500& 0.1&$-$&$-$\\
10&2MASS~J1523+3014& L8 &1700&5.5&1600&&1700&4.5& 1500& 3.0&$-$&$-$\\
\enddata
\end{deluxetable}
\end{landscape}

\section{Discussions}
\subsection{Comparison with 0.01$\mu$m best-fit model}
We show the comparison between observation and two models, the new best-fit model and the 0.01$\mu$m best-fit model, for ten L-dwarfs in Figure~\ref{figall}.
The $G_k$ value of eight of the ten objects (except for 2MASS~J2224-0158 and SDSS~1446+0024) were improved using the new best-fit model.
Among them, five objects (2MASS~1439+1929, GJ~1001B, 2MASS~J0825+2115, 2MASS~J1632+1904, and 2MASS~J1523+3014) demonstrated improvement in their reduced $\chi^2$ values for both the SpeX/CGS4 and {\AKARI} data. 
Conversely, the other three objects have worse values of reduced $\chi^2$ for either the SpeX/CGS4 or the {\AKARI} data;
the reduced $\chi^2$ value for the SpeX/CGS4 region, $\chi^2_{\mathrm{spex}}$, becomes worse for two of the three objects (SDSS~J0539--0059 and 2MASS~J1507--1627), 
while the reduced $\chi^2$ value for the {\AKARI} region, $\chi^2_{\mathrm{akari}}$, in the best-fit model is worse for the other object (2MASS~J0036+1821).
In summary, the reduced $\chi^2_{\mathrm{akari}}$ values are improved by the best-fit model for most objects.
These results of being negative, positive, or zero are shown by $-/+/0$ for $\chi^2_{\mathrm{spex},\mathrm{new}}$-$\chi^2_{\mathrm{spex},\mathrm{0.01{\mu}\mathrm{m}}}$ and $\chi^2_{\mathrm{akari},\mathrm{new}}$-$\chi^2_{\mathrm{akari},\mathrm{0.01{\mu}\mathrm{m}}}$in columns 11 and 12 of Table~\ref{modeltable}, respectively.

\subsubsection{Cases Improved by the Introduction of Different Dust Sizes}
\begin{figure}
\epsscale{1.0} 
\begin{center}
\plotone{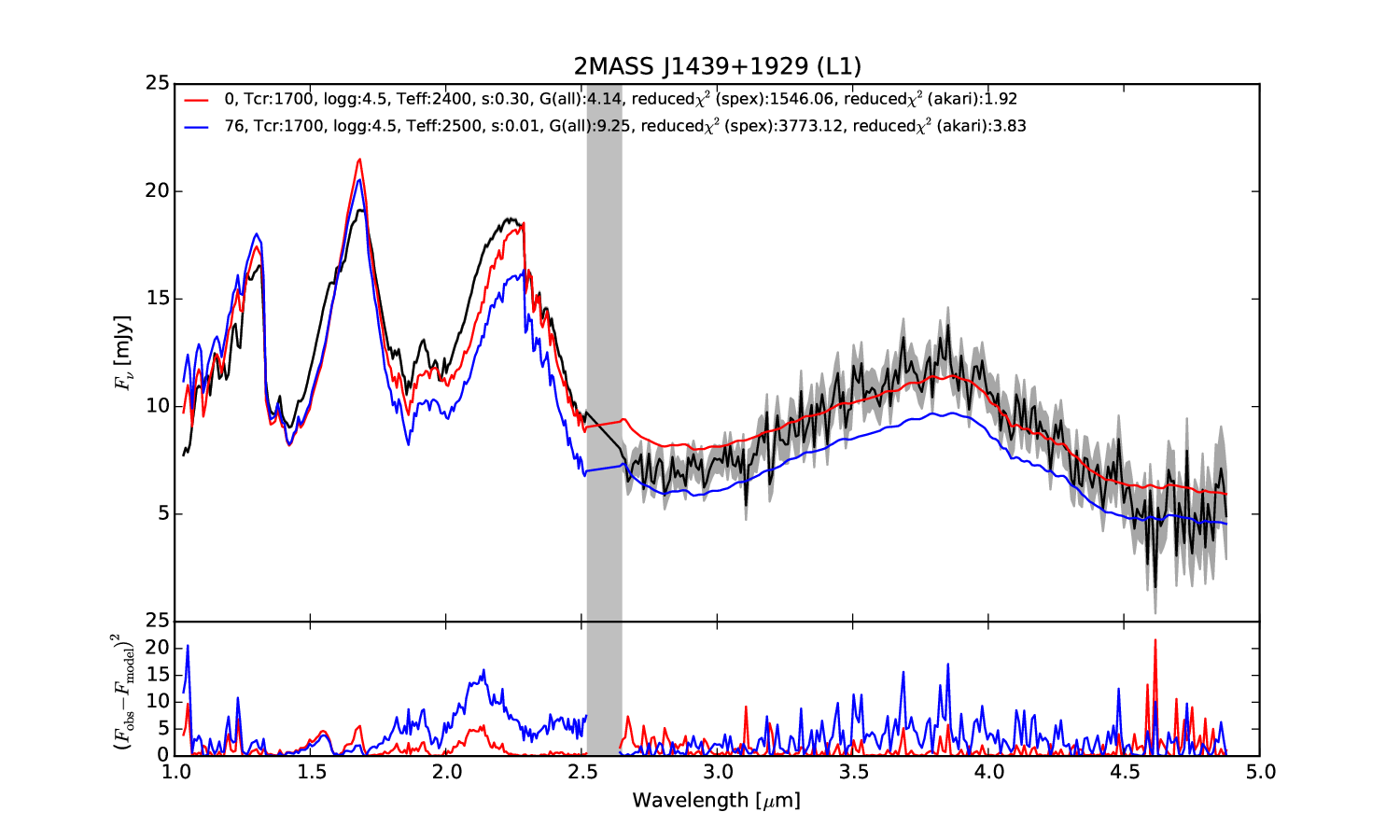}
\end{center}
\caption{Comparison between the observed spectrum and two models for 2MASS~J1439+1929 (L1). The notation is the same as in Figure~\ref{figall}.
This is an example of a significant improvement in the reproduction of the observed spectrum by model spectra due to changing the dust grain size in the model.
Bottom panel indicates the same as Figure~\ref{f9}.} 
\label{fig1}
\end{figure}

In seven of the ten objects, the new best-fit model based on changing dust sizes clearly improves the reproducibility for the observed spectrum.
For five of these seven objects that the values of both the reduced $\chi^2_{\mathrm{spex}}$ and the reduced $\chi^2_{\mathrm{akari}}$ have been improved, the new best-fit models with mid-sized dust grains ($\sim0.1-3.0$~$\mu$m) fit better than the model with a dust grain size of 0.01~$\mu$m.
In particular, 2MASS~1439+1929 is an example of an object for which the new best-fit model provides a significantly improved reproduction of the observed spectrum.
The entire {\AKARI} region and the $K$ band in the SpeX region, which do not fit with the 0.01\,{\micron} best-fit model, show an admirable improvement (Figure~\ref{fig1}).

\begin{figure}
\epsscale{1.} 
\begin{center}
\plotone{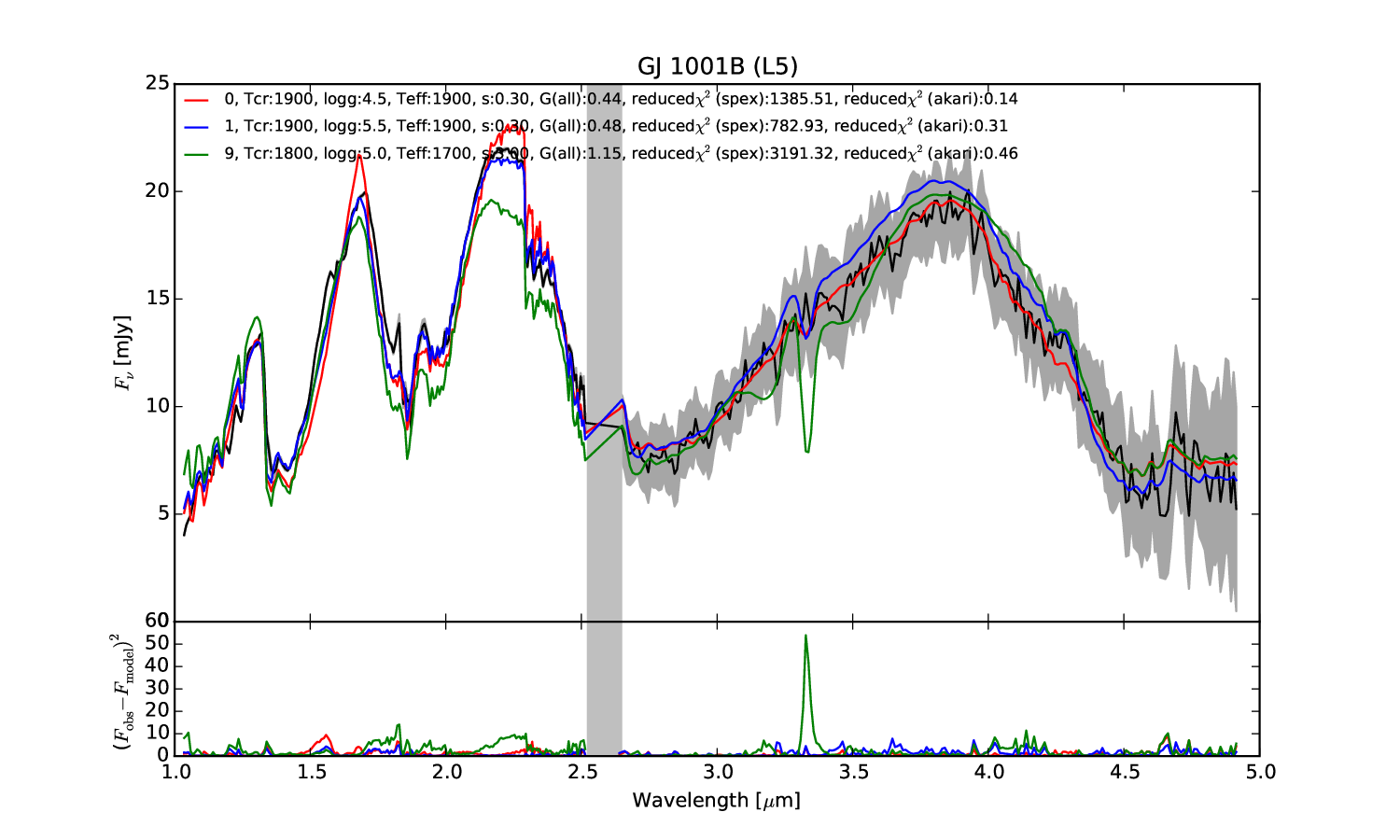}
\end{center}
\caption{Comparison between the new best-fit model (red), second model (blue), and large dust size model (green). The observed spectrum of GJ~1001B is also plotted.
Bottom panel indicates the same as Figure~\ref{f9}.} 
\label{gj1001b12}
\end{figure}

Figure~\ref{gj1001b12} shows the comparison between the spectra of GJ~1001B and the new best-fit and second best-fit models, where the values of model parameters are same except {\logg}.
The size of dust grains is 0.3 $\mu$m in both of these models.
For GJ~1001B, the spectral shape in the SpeX region is better in the second best-fit model than in the new best-fit model, but on the other hand, the spectral reproducibility in the {\AKARI} region is worse (Figure~\ref{gj1001b12}). However, the deviation of the second best-fit model from observations in the {\AKARI} region is within the range of {\AKARI} observation error. This may be due to an underestimation of the SpeX data weighting in the current analysis. 
Moreover, it is important to note that the model with $s=0.3\,\micron$ fits better to the observed spectrum than that with $s=0.01\,\micron$.

Alternatively, for 2MASS~J0036+1821 the new best-fit model with a dust size of 0.3~$\mu$m fits better than the model with dust size of 0.01~$\mu$m.
For this object, the $G_k$ value and the $\chi^2_{\mathrm{spex}}$ are improved, while the $\chi^2_{\mathrm{akari}}$ becomes slightly worse. 
However, the differences of $\chi^2_{\mathrm{akari}}$ are small between the new best-fit and 0.01 best-fit models ($\sim$4\%), while the differences between the spectral shapes of the new best-fit model and 0.01~$\mu$m best-fit model in the {\AKARI} regions are also small, lying within the error of the {\AKARI} data.
In addition, the flux value around 2.8~$\mu$m for the new best-fit model fits the observed spectrum better than that of the 0.01~$\mu$m best-fit model.
Therefore, we can infer that the new best-fit model describes the observed data more accurately for this object too.

In addition to these six objects, the best-fit model for SDSS~1446+0024, which is based on a dust size of 0.01~$\mu$m, is able to explain the entire observed spectrum in the context of the relatively poor signal-to-noise ratio in the {\AKARI} region.

From these results, 
it is suggested that the dust size in the atmosphere of the majority of brown dwarfs is neither especially small nor especially large, and that dust particles with a diameters spanning 0.1--3.0~$\mu$m play an important role in determining atmospheric structure-related properties of real brown dwarfs, albeit with some exceptions, e.g. SDSS~1446+0024.
Furthermore, the results reveal that there is no correlation between dust size and effective temperature {\Teff} or mass {\logg}.

\subsubsection{Cases not Improved by the Introduction of Different Dust Sizes}
On the other hand, for two of three objects yet to be discussed (SDSS~0539--0059 and 2MASS~J1507--1627), the new best-fit model produced improvements in the $G_k$ values and the reduced $\chi^2_{\mathrm{akari}}$, while the reduced $\chi^2_{\mathrm{spex}}$ worsened (Figure~\ref{figall}d and \ref{figall}f). 
In these cases, owing to the large change in the reduced $\chi^2_{\mathrm{spex}}$ (150--300\%),
and the fact that the shape of the model spectra deviates significantly from the observed spectra, the model reproducibility with respect to the observed data was not satisfactory, despite the improvement in the $G_k$ values.
However, it should be noted that the new best-fit model reproduces the {\AKARI} region in the observed spectra with greater success than the 0.01~$\mu$m best-fit model.
For the last object, 2MASS~J2224-0158, the best-fit model corresponded to a dust size of 0.01~$\mu$m, a large deviation are observed within the $K$ band, and no model was able to explain this $K$ band in the observed spectrum.

For these three objects,
it is possible that alternative treatments need to be considered, such as the differences in chromospheric activity (\citealt{Sorahana_2014a}) and/or elemental abundances (\citealt{Sorahana_2014b}), alongside dust grain size.
Indeed, accounting for chromospheric activity has been shown to reproduce the observed spectra favorably in models of the 2MASS~J2224-0158 spectrum (\citealt{Sorahana_2014a}).

\subsection{Comparison with other dust size models}
\begin{figure}
\epsscale{1.} 
\begin{center}
\plotone{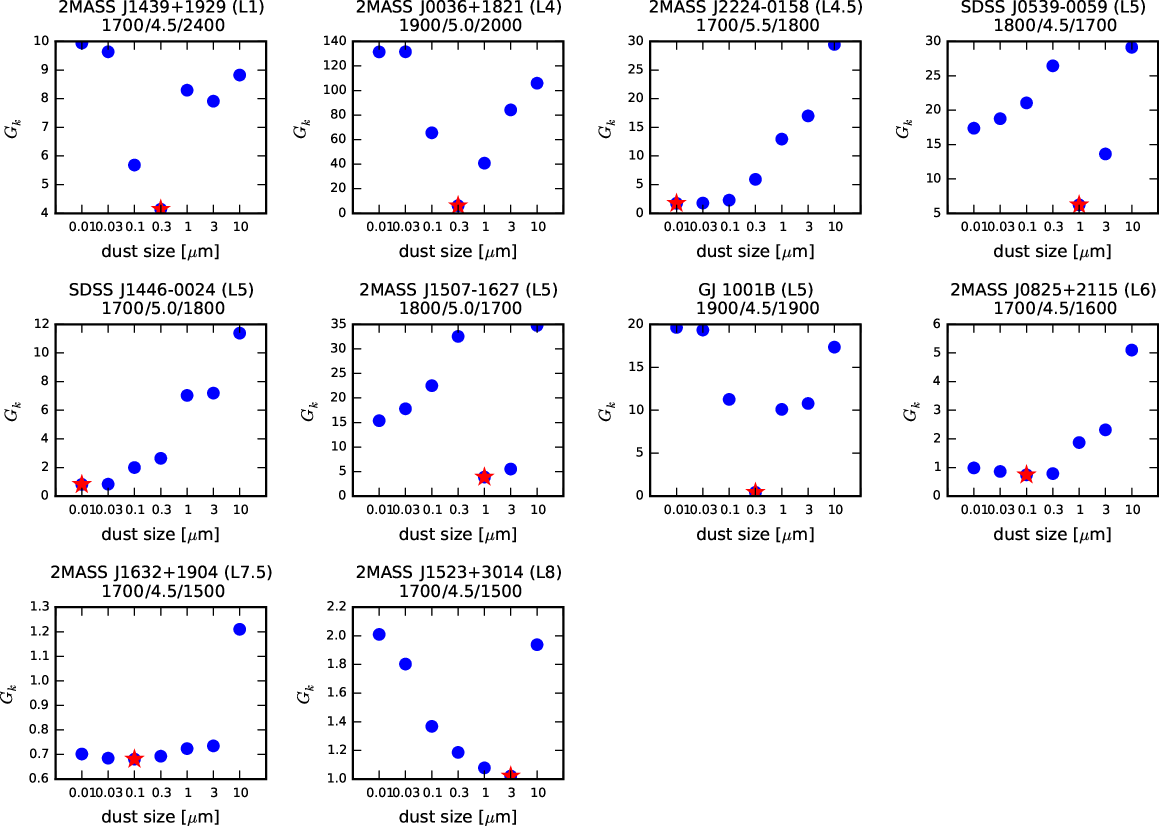}
\end{center}
\caption{$G_k$ map of derived best-fit model for each {\AKARI} object. The stars represent the $G_k$ values of the new best-fit models, and the circles indicate the $G_k$ values of the other dust size models with otherwise identical parameter sets to the new best-fit model.
} 
\label{chi2map}
\end{figure}

In order to distil the effects of dust size alone, we compared the best-fit model to other dust size models while retaining fixed values for all other model parameters ({\Tcr}, {\logg}, and {\Teff}).
The $G_k$ map of the derived best-fit model for each object is shown in Figure~\ref{chi2map}.
These results show that the dust size can be well constrained in all but one of the objects included in this study, namely 2MASS~J1632+1904.
This finding indicates strongly that dust size is an important parameter in most cases.
By contrast, object 2MASS~J1632+1904 has a degenerate dust size, and there is not much difference in the $G_k$ values, except for the $s=10\,\micron$ models.
Similarly, the 2MASS~J2224--0158 and 2MASS~J0825+2115 do not differ much in $G_k$ value between the small and middle size models.
For these two objects, we are able to refine the effect of dust size by actually comparing the spectra of each model with each other, as we did in Sections 5.2 and 6.1. 

\subsection{Comparison with other models with relatively small $G_k$ values.}
We investigated the extent to which the various models actually reproduced the shape of the observed spectra.
We compared the dust size of the best-fit model to other models with relatively small $G_k$ values, 
and investigated the reproducibility of models based on large dust sizes ($s\ge3.0\,\micron$).
The results of the comparison for each object are summarized below.

\paragraph*{2MASS~J1439+1929}
Most of the models with small values of $G_k$ have $s=0.3\,\micron$.
Models using small and large dust sizes do not rank higher in terms of $G_k$ values.
Therefore, in addition to the other parameters ({\Tcr}, {\logg}, and {\Teff}), the dust size is very sensitive for this object.

\paragraph*{2MASS~J0036+1821}
In addition to the mid-sized (0.1--3.0~$\mu$m) dust models, there is also a small ($s\le0.03\,\micron$) dust size model that yielded a small $G_k$ value. 
However, smaller dust size models corresponded to poorer overall reproducibility of the observed data in the SpeX region relative to medium dust size models, as shown in Figure~\ref{figall}b. Conversely, no large ($s \ge3.0\,\micron$) dust size model produced a small $G_k$ value.

\begin{figure}
\epsscale{1.} 
\begin{center}
\plotone{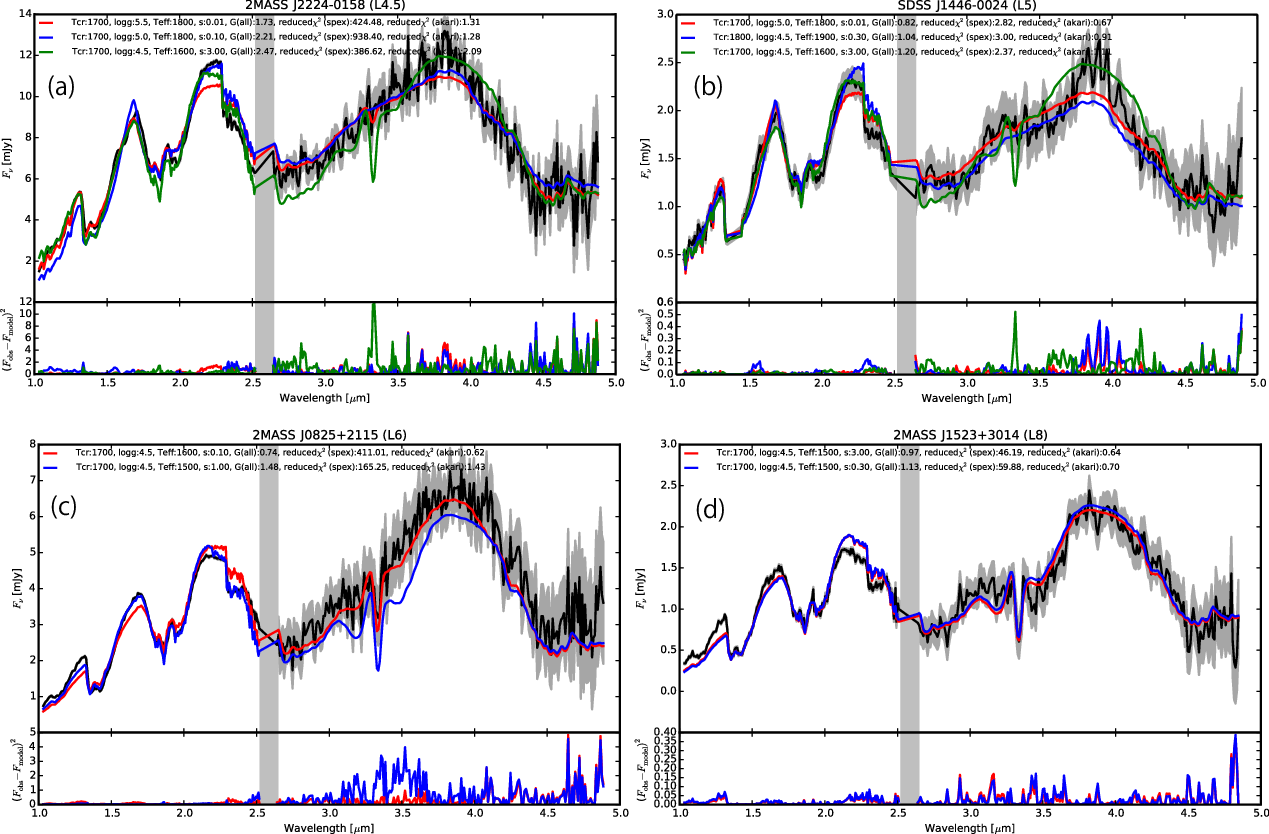}
\end{center}
\caption{Comparison between the observed spectrum and a selection of modeled spectra for four objects. Bottom panel of each Figure indicates the same as Figure~\ref{f9}.
(a) Comparison with models based on a small dust size of 0.01\,\micron (red), medium dust size of 0.1\,\micron (blue), and large dust size of 3\,\micron (green) with small $G_k$ for 2MASS~J2224--0158. 
(b) Comparison with models based on a small dust size of 0.01\,\micron (red), medium dust size of 0.3\,\micron (blue), and large dust size of 3\,\micron (green) with small $G_k$ for SDSS~J1446+0024.
(c) Comparison with models based on medium dust sizes of 0.1 (red) and 1.0~$\mu$m (blue) with small $G_k$ for 2MASS~J0825+2115.
(d) Comparison with models based on a medium dust size of 0.3~$\mu$m (blue) and large dust size of 3.0~$\mu$m (red) with small $G_k$ for 2MASS~J1523+3014.} 
\label{F1}
\end{figure}

\paragraph*{2MASS~J2224-0158}
Little difference was observed between the best performing models with small $G_k$ values. In addition to $s=0.01\mu$m, some of these models have $s\ge0.1\mu$m 
At $\lambda<1.8\mu$m, the models with $s=0.01\mu$m and $0.03\mu$m provide closer agreement with the observed data, but $s=0.1\mu$m provides superior reproducibility for $1.8<\lambda<2.3\mu$m.
Therefore, both small and medium dust grain sizes are justifiable candidates for the best-fit model.
Figure~\ref{F1} shows the comparison between the observations of several objects and  models with the small dust size  ($\sim 0.01\,\mu$m) and/or medium dust size ($\sim 0.1\,\mu$m) and/or large dust size ($\sim 3\,\mu$m). 
Although there are a few large dust size models among the small $G_k$ value models, these models were unable to reproduce the {\AKARI} region of the spectrum sufficiently, as shown in Figure~\ref{F1}a. Therefore, the large dust size models were discounted from the fitting models.

\paragraph*{SDSS~J0539--0059 and 2MASS~J1507--1627}
Most of the best performing models were not able to provide complementary reproducibility for both the SpeX and {\AKARI} regions. If the model agreed with the observation in the SpeX regime, the spectral shape in the {\AKARI} regime deviated from the observation, 
and vice versa as shown in Figure~\ref{figall}d and \ref{figall}f.
Therefore, it also proved difficult to judge the quality of the observed data reproduction by eye, even for the best performing models with small $G_k$ values.

\paragraph*{SDSS~J1446+0024}
The best performing models were dominated by models with relatively small dust sizes, although a few possessed medium dust sizes. 
In the cases of medium dust size models with relatively small $G_k$ values, 
the spectrum are partially poorly fitted to the observed spectrum, e.g., the slope at 1.5 $\mu$m, CO absorption band at 2.3 $\mu$m, and the peak of L band deviate from the observed data.
For models with {\Teff}=1600~K, the $G_k$ value of the models with large dust sizes ($\sim1$ or $3\,\micron$) were small. 
However, the spectral shape for the {\AKARI} region, especially the methane band around 3.3~$\mu$m, displayed deviations between the simulated and observed spectra (Figure~\ref{F1}b).
Models with small dust sizes outperformed the other models comprehensively with respect to fitting the observed data. 

\paragraph*{GJ~1001B}
The best performing models those employing relatively small to medium dust sizes. 
However, for the models with small dust sizes, the peak of the $L$ band in the {\AKARI} region did not provide a satisfactory fit, as shown in Figure~\ref{figall}g.
By contrast, the models with large dust sizes did not fit the spectra in the SpeX regime or at the 3.3~$\mu$m {\CHf} absorption band, as shown in Figure~\ref{gj1001b12}.

\paragraph*{2MASS~J0825+2115}
The best performing models were those for which $s<0.3\mu$m.
However, in comparison to the small dust size models, the middle dust size model provided a more comprehensive fit as shown in Figure~\ref{figall}h, although the difference is small.
The models with $s\ge1.0\mu$m did not produce a comparable fitting performance because the {\AKARI} region deviated from the observed data, even though the reproducibility of the SpeX region was adequate (Figure~\ref{F1}c).
Therefore, in considering the entire wavelength region, we were able to dismiss the models with large dust sizes.

\paragraph*{2MASS~J1632+1904}
Comparing the best performing models, the absence of large dust size models, e.g. $s\ge1.0\mu$m, was notable. Furthermore, little difference was observed between the smaller dust size models and the middle size models, making it difficult to decide between them because of the low signal-to-noise ratio of the {\AKARI} data.

\paragraph*{2MASS~J1523+3014}
Little difference was observed between the smaller dust size models and the middle size models as observed for 2MASS~J1632+1904, although the middle dust size model provided a closer agreement to the observed spectrum in comparison with the best-fit model of 0.01~$\mu$m, as shown in Figure~\ref{figall}j. 
Although the dust size of best-fit model is relatively large, 3~$\mu$m, the middle dust size models are applicable because the spectral shapes of the middle dust size models from $0.1<s\le3\,\micron$ show minimal differences (Figure~\ref{F1}d). 
There were no larger dust size models (10~$\mu$m) among the best performing models. 

From these results, it is evident that models with middle dust sizes are more reproducible to the observed spectra than models with smaller or larger dust sizes, 
and models with large dust sizes can be dismissed in the case of most objects.

\subsection{The Deviations between Observational and Model Spectra}
Our demonstration that the spectra of most objects are improved by a new best-fit model based on medium dust sizes (0.1--3.0~$\mu$m) is consistent with other brown dwarf atmospheric models, such as \citet{Allard_2003}, \citet{Ackerman_2001}, and \citet{Helling_2008c}, which suggested that the dust could grow up to a few $\mu$m after nucleation.

For most objects, however, even models with larger dust sizes than 0.01~$\mu$m, the particle diameter applied in the previous UCM, do not perfectly reproduce the shape of the spectrum.
In our model, we assume a singly value of dust size in the atmosphere. 
However, it is possible that the reproduction of the observed spectra could be improved further by considering the dust size distribution in the altitude direction. 
Actually the dust is expected to grow in size and settle when it reaches a certain size in the atmosphere.

The photospheric surface in the brown dwarf atmospheres differs for each wavelength.
As shown in Section~4, the photospheric surface ($\tau_\lambda\sim1$) of the gas-dominated regime corresponds to the relatively high altitude in the atmosphere, while the photospheric surface of the dust-dominated regime corresponds to the lower altitude.  
As can be seen from the spectral changes due to dust size variation in Figure~\ref{f9}, in the case of the parameter set $(${\Tcr}/{\logg}/{\Teff}$)=(1700$K$/4.5/1600$K), 
the model with a large dust size of 3.0~$\mu$m fits better for the $J$ and $H$ bands.
In this case, the photospheric surface ($\tau_\lambda\sim1$) for the $J$ and $H$ bands is close to the tops of the clouds, as shown by the red circle in Figure~\ref{qext}, e.g. the dust opacity of $s=3$~$\mu$m is close to 0.2 cm$^2$/g.
By contrast, the models with small to medium dust sizes provide closer fits for the $K$ and $L$ bands, 
and the photoshperic surface of the $K$ and $L$ bands is in the cloud interior, e.g. the dust opacities of $s=0.01-0.1$~$\mu$m are well below 0.2 cm$^2$/g.
In other words, the observed spectra of 2MASS~J0825+2115 may indicate that the dust size increases from the cloud bottom to the cloud top.
This suggests that the distribution of dust grain size as a function of altitude ought to be considered.

Another factor that may influence atmospheric dynamics, such as internal turbulence, in brown dwarfs is the extent to which dust can grow and float without sedimentation. 
Therefore, if we can determine the distribution of dust size from observations, we may be able to constrain the internal turbulence.
For example, the atmospheres of SDSS~J2224--0158 (L4.5) and SDSS~J1446+0024 (L5), which correspond to best-fit models with a small dust size (0.01~$\mu$m), may have few dynamic processes, resulting in a relatively small of dust sizes.
Consequently, determining the size distributions of dust from observations and limiting the dynamic processes in the atmosphere will provide the focus of our future work.

\section{Summary}
In this study, we focused on the size of dust grains in brown dwarf atmospheres, investigating how the opacity is affected as the dust size changes and, ultimately, how this interplay impacts our ability to reproduce the infrared spectra of brown dwarfs accurately via simulations.

We confirmed that the change in dust size affects the dust opacity and, thus, the temperature-pressure structure of the atmosphere, 
leading to variations in the shape of the spectrum.
We considered the effect on the spectral shape in two regimes: the dust- and gas-dominated regimes, in which dust and gas are the dominant absorbing elements, respectively.
The flux is roughly determined by the blackbody radiation of the photosphere for each $\lambda$ and depends on the photospheric temperature at $\tau_\lambda\sim1$. 
We found that the photosphere for the dust-dominated regime exists around the dust layer, with the photospheric surface depending on the dust opacity.
A large dust opacity results in a low photospheric temperature
because the photospheric surface is close to the cloud top, where the temperature in the dust layers is relatively lower (the temperature increases moving deeper into the atmosphere for brown dwarfs).
Therefore, the flux value is also small for the case of large dust opacity for the dust-dominated regime, and we infer that the dust opacity affects the flux value directly. 

Conversely, for the gas-dominated regime, a large dust opacity results in high photospheric temperatures, which creates large flux emissions. This is because the temperature does not decrease easily before it reaches the photosphere when the dust opacity is large. 
Thus, in this instance, the dust opacity affects the amount of flux indirectly.

In addition, we attempted to constrain the predominant dust size in the atmospheres of 10 objects by comparing their observed spectra to 1051 models, for which the additional parameter of dust size $s$ was adopted.
For six of the ten L dwarfs we examined, the observed data could be reproduced more completely by a model with a dust size other than $s=0.01\,\micron$, with the dust size of our new best-fit models ranging from $s=0.1$--3.0~$\mu$m, representing medium grain sizes.
In addition to these six objects, the best-fit model for SDSS~1446+0024, which was based on a dust size of 0.01~$\mu$m, was able to explain the entire observed spectrum in the context of the relatively poor signal-to-noise ratio in the {\AKARI} region, but this was an exceptional case.

We also showed that some new-best fit models provided closer agreement with the observed spectra than original models, which are based on a dust size of 0.01~$\mu$m, for most objects, although perfect correspondence between observation and model remained elusive.
Therefore we will also consider the distribution of dust sizes within different atmospheric layers as a function of altitude in future studies.

For the remaining three objects, no models provided adequate reconstructions of the observed spectra. 
Other treatments, such as the differences in chromospheric activity and/or elemental abundances, in addition to dust size may be required to improve the fidelity of spectra reproductions in these cases. These considerations will be explored in our future work.

We acknowledge Dr. Adam Burgasser, Dr. Sandy Leggett and Dr. Michael. Cushing for providing access to the observed near-infrared spectral data.
This research is based on observations with {\AKARI}, a JAXA project with the participation of the ESA. 
Also, we thank Prof. Takashi Tsuji for kindly granting us access to the UCM and for helpful suggestions. 
This work is supported in part by JSPS KAKENHI grant Nos. 18K03689 and 20H05657. 
H.K. acknowledges support from the JSPS KAKENHI grant Nos of 20H04612, 18H05436, 18H05438, 17H01105, 17K05632, and 17H01103.




\bibliography{sorahana}

\begin{thebibliography}{46}
\expandafter\ifx\csname natexlab\endcsname\relax\def\natexlab#1{#1}\fi

\bibitem[{{Ackerman} \& {Marley}(2001)}]{Ackerman_2001}
{Ackerman}, A.~S., \& {Marley}, M.~S. 2001, ApJ, 556, 872

\bibitem[{{Allard} {et~al.}(2001){Allard}, {Hauschildt}, {Alexander},
  {Tamanai}, \& {Schweitzer}}]{Allard_2001a}
{Allard}, F., {Hauschildt}, P.~H., {Alexander}, D.~R., {Tamanai}, A., \&
  {Schweitzer}, A. 2001, ApJ, 556, 357

\bibitem[{{Allard} {et~al.}(2003){Allard}, {Allard}, {Hauschildt}, {Kielkopf},
  \& {Machin}}]{Allard_2003}
{Allard}, N.~F., {Allard}, F., {Hauschildt}, P.~H., {Kielkopf}, J.~F., \&
  {Machin}, L. 2003, A\&A, 411, L473

\bibitem[{{Allende Prieto} {et~al.}(2002){Allende Prieto}, {Lambert}, \&
  {Asplund}}]{Allende_2002}
{Allende Prieto}, C., {Lambert}, D.~L., \& {Asplund}, M. 2002, ApJ, 573, L137

\bibitem[{{Burgasser}(2007)}]{Burgasser_2007}
{Burgasser}, A.~J. 2007, ApJ, 659, 655

\bibitem[{{Burgasser} {et~al.}(2010){Burgasser}, {Cruz}, {Cushing}, {Gelino},
  {Looper}, {et~al.}}]{Burgasser_2010}
{Burgasser}, A.~J., {Cruz}, K.~L., {Cushing}, M., {et~al.} 2010, ApJ, 710, 1142

\bibitem[{{Burgasser} {et~al.}(2006){Burgasser}, {Geballe}, {Leggett},
  {Kirkpatrick}, \& {Golimowski}}]{Burgasser_2006a}
{Burgasser}, A.~J., {Geballe}, T.~R., {Leggett}, S.~K., {Kirkpatrick}, J.~D.,
  \& {Golimowski}, D.~A. 2006, ApJ, 637, 1067

\bibitem[{{Burgasser} {et~al.}(2002){Burgasser}, {Kirkpatrick}, {Brown},
  {Reid}, {Burrows}, {et~al.}}]{Burgasser_2002}
{Burgasser}, A.~J., {Kirkpatrick}, J.~D., {Brown}, M.~E., {et~al.} 2002, ApJ,
  564, 421

\bibitem[{{Burgasser} {et~al.}(2008){Burgasser}, {Liu}, {Ireland}, {Cruz}, \&
  {Dupuy}}]{Burgasser_2008}
{Burgasser}, A.~J., {Liu}, M.~C., {Ireland}, M.~J., {Cruz}, K.~L., \& {Dupuy},
  T.~J. 2008, ApJ, 681, 579

\bibitem[{{Burgasser} {et~al.}(2004){Burgasser}, {McElwain}, {Kirkpatrick},
  {Cruz}, {Tinney}, \& {Reid}}]{Burgasser_2004}
{Burgasser}, A.~J., {McElwain}, M.~W., {Kirkpatrick}, J.~D., {et~al.} 2004, AJ,
  127, 2856

\bibitem[{{Burrows} {et~al.}(2001){Burrows}, {Hubbard}, {Lunine}, \&
  {Liebert}}]{Burrows_2001}
{Burrows}, A., {Hubbard}, W.~B., {Lunine}, J.~I., \& {Liebert}, J. 2001, RvMP,
  73, 719

\bibitem[{{Burrows} \& {Sharp}(1999)}]{Burrows_1999}
{Burrows}, A., \& {Sharp}, C.~M. 1999, \apj, 512, 843

\bibitem[{{Burrows} {et~al.}(2006){Burrows}, {Sudarsky}, \&
  {Hubeny}}]{Burrows_2006}
{Burrows}, A., {Sudarsky}, D., \& {Hubeny}, I. 2006, \apj, 640, 1063

\bibitem[{Chackerian \& Tipping(1983)}]{Chackerian_1983}
Chackerian, C.~J., \& Tipping, R.~H. 1983, J. Mol. Spectrosc., 99, 431

\bibitem[{{Charnay} {et~al.}(2018){Charnay}, {B{\'e}zard}, {Baudino},
  {Bonnefoy}, {Boccaletti}, \& {Galicher}}]{Charnay_2018}
{Charnay}, B., {B{\'e}zard}, B., {Baudino}, J.~L., {et~al.} 2018, \apj, 854,
  172

\bibitem[{{Cooper} {et~al.}(2003){Cooper}, {Sudarsky}, {Milsom}, {Lunine}, \&
  {Burrows}}]{Cooper_2003}
{Cooper}, C.~S., {Sudarsky}, D., {Milsom}, J.~A., {Lunine}, J.~I., \&
  {Burrows}, A. 2003, ApJ, 586, 1320

\bibitem[{{Cushing} {et~al.}(2008){Cushing}, {Marley}, {Saumon}, {Kelly},
  {Vacca}, {et~al.}}]{Cushing_2008}
{Cushing}, M.~C., {Marley}, M.~S., {Saumon}, D., {et~al.} 2008, ApJ, 678, 1372

\bibitem[{{Cushing} {et~al.}(2004){Cushing}, {Vacca}, \&
  {Rayner}}]{Cushing_2004}
{Cushing}, M.~C., {Vacca}, W.~D., \& {Rayner}, J.~T. 2004, PASP, 116, 362

\bibitem[{Freedman {et~al.}(2008)Freedman, Marley, \& Lodders}]{Freedman_2008}
Freedman, R.~S., Marley, M.~S., \& Lodders, K. 2008, ApJS, 174, 504

\bibitem[{{Geballe} {et~al.}(2002){Geballe}, {Knapp}, {Leggett}, {Fan},
  {Golimowski}, {et~al.}}]{Geballe_2002}
{Geballe}, T.~R., {Knapp}, G.~R., {Leggett}, S.~K., {et~al.} 2002, ApJ, 564,
  466

\bibitem[{Guelachivili {et~al.}(1983)Guelachivili, De~Villeneuve, Farrenq,
  Urban, \& Verges}]{Guelachvili_1983}
Guelachivili, G., De~Villeneuve, D., Farrenq, R., Urban, W., \& Verges, J.
  1983, J. Mol. Spectrosc., 98, 64

\bibitem[{{Helling} {et~al.}(2001){Helling}, {Oevermann}, {L{\"u}ttke},
  {Klein}, \& {Sedlmayr}}]{Helling_2001a}
{Helling}, C., {Oevermann}, M., {L{\"u}ttke}, M.~J.~H., {Klein}, R., \&
  {Sedlmayr}, E. 2001, A\&A, 376, 194

\bibitem[{{Helling} {et~al.}(2008){Helling}, {Woitke}, \&
  {Thi}}]{Helling_2008c}
{Helling}, C., {Woitke}, P., \& {Thi}, W.~F. 2008, \aap, 485, 547

\bibitem[{{Kawada} {et~al.}(2007){Kawada}, {Baba}, {Barthel}, {Clements},
  {Cohen}, {et~al.}}]{Kawada_2007}
{Kawada}, M., {Baba}, H., {Barthel}, P.~D., {et~al.} 2007, PASJ, 59, 389

\bibitem[{{Kirkpatrick} {et~al.}(2000){Kirkpatrick}, {Reid}, {Liebert},
  {Gizis}, {Burgasser}, {et~al.}}]{Kirkpatrick_2000}
{Kirkpatrick}, J.~D., {Reid}, I.~N., {Liebert}, J., {et~al.} 2000, AJ., 120,
  447

\bibitem[{{Larimer}(1967)}]{Larimer_1967a}
{Larimer}, J.~W. 1967, GeCoA, 31, 1215

\bibitem[{{Larimer} \& {Anders}(1967)}]{Larimer_1967b}
{Larimer}, J.~W., \& {Anders}, E. 1967, GeCoA, 31, 1239

\bibitem[{{Lord}(1965)}]{Lord_1965}
{Lord}, III, H.~C. 1965, lcar, 4, 279

\bibitem[{{Marley} {et~al.}(2002){Marley}, {Seager}, {Saumon}, {Lodders},
  {Ackerman}, {et~al.}}]{Marley_2002}
{Marley}, M.~S., {Seager}, S., {Saumon}, D., {et~al.} 2002, ApJ, 568, 335

\bibitem[{{Morley} {et~al.}(2012){Morley}, {Fortney}, {Marley}, {Visscher},
  {Saumon}, \& {Leggett}}]{Morley_2012}
{Morley}, C.~V., {Fortney}, J.~J., {Marley}, M.~S., {et~al.} 2012, \apj, 756,
  172

\bibitem[{{Nakajima} {et~al.}(2001){Nakajima}, {Tsuji}, \&
  {Yanagisawa}}]{Nakajima_2001}
{Nakajima}, T., {Tsuji}, T., \& {Yanagisawa}, K. 2001, ApJ, 561, L119

\bibitem[{{Onaka} {et~al.}(2007){Onaka}, {Matsuhara}, {Wada}, {Fujishiro},
  {Fujiwara}, {et~al.}}]{Onaka_2007}
{Onaka}, T., {Matsuhara}, H., {Wada}, T., {et~al.} 2007, PASJ, 59, 401

\bibitem[{Partridge \& Schwenke(1997)}]{Partridge_1997}
Partridge, H., \& Schwenke, D.~W. 1997, J. Chem. Phys., 106, 4618

\bibitem[{{Rothman}(1997)}]{Rothman_1997}
{Rothman}, L.~S. 1997, {High-temperature Molecular Spectroscopic Database
  (CD-ROM)} (Andover: ONTAR Co.)

\bibitem[{{Saumon} \& {Marley}(2008)}]{Saumon_2008}
{Saumon}, D., \& {Marley}, M.~S. 2008, ApJ, 689, 1327

\bibitem[{{Sorahana} {et~al.}(2014){Sorahana}, {Suzuki}, \&
  {Yamamura}}]{Sorahana_2014a}
{Sorahana}, S., {Suzuki}, T.~K., \& {Yamamura}, I. 2014, \mnras, 440, 3675

\bibitem[{{Sorahana} \& {Yamamura}(2012)}]{Sorahana_2012}
{Sorahana}, S., \& {Yamamura}, I. 2012, \apj, 760, 151

\bibitem[{{Sorahana} \& {Yamamura}(2014)}]{Sorahana_2014b}
---. 2014, \apj, 793, 47

\bibitem[{{Tremblin} {et~al.}(2016){Tremblin}, {Amundsen}, {Chabrier},
  {Baraffe}, {Drummond}, {Hinkley}, {Mourier}, \& {Venot}}]{Tremblin_2016}
{Tremblin}, P., {Amundsen}, D.~S., {Chabrier}, G., {et~al.} 2016, \apjl, 817,
  L19

\bibitem[{{Tsuji}(2002)}]{Tsuji_2002}
{Tsuji}, T. 2002, ApJ, 575, 264

\bibitem[{{Tsuji}(2005)}]{Tsuji_2005}
---. 2005, ApJ, 621, 1033

\bibitem[{{Tsuji} {et~al.}(1996{\natexlab{a}}){Tsuji}, {Ohnaka}, \&
  {Aoki}}]{Tsuji_1996a}
{Tsuji}, T., {Ohnaka}, K., \& {Aoki}, W. 1996{\natexlab{a}}, A\&A, 305, L1

\bibitem[{{Tsuji} {et~al.}(1996{\natexlab{b}}){Tsuji}, {Ohnaka}, {Aoki}, \&
  {Nakajima}}]{Tsuji_1996b}
{Tsuji}, T., {Ohnaka}, K., {Aoki}, W., \& {Nakajima}, T. 1996{\natexlab{b}},
  A\&A, 308, L29

\bibitem[{{Wenger} \& {Champion}(1998)}]{Wenger_1998}
{Wenger}, C., \& {Champion}, J.~P. 1998, \jqsrt, 59, 471

\bibitem[{{Woitke} \& {Helling}(2003)}]{Woitke_2003}
{Woitke}, P., \& {Helling}, C. 2003, A\&A, 399, 297

\bibitem[{{Woitke} \& {Helling}(2004)}]{Woitke_2004}
---. 2004, A\&A, 414, 335

\end{thebibliography}

\end{document}